\documentclass[11pt,twoside,a4paper]{article}

\usepackage{Sweave}
\usepackage{amssymb,amsbsy,amsmath,amsfonts,amssymb,amscd, amsthm}
\usepackage{mathrsfs}
\usepackage{color, epsfig, graphicx}
\usepackage{subfigure}
\usepackage{anysize} 
\usepackage{natbib}
\usepackage{url}

\marginsize{2.5cm}{2.5cm}{2cm}{4cm}
\newcommand{\summ}[2]{\sum _{#1}^{#2}} 
\newcommand{\prodd}[2]{\prod _{#1}^{#2}} 
\newcommand{\bs}{\boldsymbol} 
\newcommand{\vb}{\texttt} 

\title{Extending INLA to a class of near-Gaussian latent models}
\author{Thiago G. Martins and H\aa vard Rue\\
Department of Mathematical Sciences\\ 
Norwegian University of Science and Technology\\
N-7491 Trondheim, Norway}
\date{\today}

\begin{document}
\Sconcordance{concordance:inla_ngnodes_revised.tex:inla_ngnodes_revised.Rnw:%
1 480 1 1 14 642 1}

\maketitle

\begin{abstract}
This work extends the Integrated Nested Laplace Approximation (INLA) method to latent models 
outside the scope of latent Gaussian models, where independent components of the latent field can 
have a near-Gaussian distribution. The proposed methodology is an essential component of a 
bigger project that aim to extend the \texttt{R} package \texttt{INLA} (\texttt{R-INLA}) 
in order to allow the user to add flexibility and challenge the Gaussian assumptions 
of some of the model components in a straightforward and intuitive way. Our approach is 
applied to two examples and the results are compared with that obtained by Markov Chain Monte 
Carlo (MCMC), showing similar accuracy with only a small fraction of computational time. 
Implementation of the proposed extension is available in the \texttt{R-INLA} package.
\end{abstract}


\section{Introduction}\label{sec:intro}

Integrated Nested Laplace Approximation (INLA) is an approach proposed by 
\cite{rue2009approximate} to perform approximate fully Bayesian inference on the class
of latent Gaussian models (LGM). It was demonstrated in the original paper that, when 
compared with the more usual Markov Chain Monte Carlo (MCMC) schemes 
\citep{robert2004monte, gamerman2006markov}, INLA
outperforms the latter both in terms of accuracy and speed.
Monte Carlo averages under MCMC are characterized by additive $O_p(N^{-1/2})$ errors, where
$N$ is the simulated sample size, meaning that we need $100$ times more computational
time to improve our estimates by one digit. Besides that, due to the additive
nature of Monte Carlo estimates, it is even harder to accurately estimate 
tail probabilities with MCMC. On the other hand, INLA bypasses the need for stochastic 
simulation by an extensive use of simple and fast Gaussian approximations to take 
advantage of the properties of latent Gaussian models, where for most real problems 
and data sets, the conditional posterior of the latent field is typically well behaved, 
being close to a Gaussian. As opposed to MCMC, INLA has relative error which allow for more
accurate estimates of small quantities, as for example the estimation of tail probabilities.

INLA is not meant to be a replacement of MCMC in applied statistics, but
it is a specific tailored algorithm that works extremely well in the broad
class of LGMs, and thus offers a better option in this
context. 
Given the wide range of models that belong to the class of LGMs, it is common to
start our analysis with some standard model $M_0$, contained within the LGM class,
to solve a particular set of problems.
However, we sometimes realize that our analysis would be compromised if the data has 
specific deviations from the modeling assumptions in $M_0$.
A natural way to robustify $M_0$ against these deviations is to embed it into a larger,
more flexible model $M_1$ \citep{box1980sampling}, possibly outside the realm of LGMs.
To give a more specific example, assume we start with a simple linear mixed effects model
\citep{laird1982random} with a Gaussian random effect. It is well known in the literature 
that the Gaussian distribution
is not robust against outliers \citep{lange1989robust}, and a sensible approach would
be to embed our model into a larger model that is robust to outliers. One example is the model proposed by \cite{pinheiro2001efficient}, where Student's t distributions are used for both the error
terms and the random effects. Unfortunately, although the current implementation of INLA can 
deal with non-Gaussian likelihoods, it cannot handle non-Gaussian latent components, as the
Student's t random effects in this example.

This paper proposes an extension that allows INLA to be applied to models
where some independent components of the latent field have a non-Gaussian distribution.
Interest for such models arise often in 
the literature but the lack of
user friendly software able to handle them in a fast and accurate way might 
lead someone to stay with more standard models, even though it might not be the best for their 
applications. 
This proposed extension is an essential component of a 
bigger project that aim to extend the \texttt{R} package \texttt{INLA} 
(\texttt{R-INLA}) in order to allow the user to add flexibility and to 
challenge the Gaussian assumptions of some of the model components in 
a straightforward and intuitive way. Specifically, we plan to give an option
in \texttt{R-INLA} for the user that start with a initial model $M_0$ based on Gaussian 
distributions to expand the model in certain directions by adding flexibility around these
Gaussian distributions, e.g. correcting it for asymmetry and/or kurtosis.
We refer to these distributions that add flexibility around the Gaussian
as \textit{near-Gaussian distributions}, as will be explained in Section \ref{sec:INLAext:models}.
But for this
to be accomplished we need the methodology presented in this paper to be able to perform
inference in these near-Gaussian latent models.

Besides being an essential component of the model expansion feature described above, the
approach presented here is useful in itself, since it allows us to fit models that are
currently in demand and lie outside the LGM framework. Examples of such models are survival models with gamma frailty \citep{ibrahim2001bayesian} where the random 
effects have a log gamma distribution, the already mentioned robust mixed effect models 
\citep{pinheiro2001efficient}, where the distribution 
of the random effects is assumed to be Student's t rather than Gaussian, thus allowing for 
outlier identification and accommodation and non-Gaussian state space models 
\citep{kitagawa1987non}, where the distribution of the noise in the state space evolution
equation is assumed to be non-Gaussian. The list is, of course, much longer than that, 
and a reliable method to fit this large class of models efficiently will provide the 
right tools for the applied scientist to practice a more flexible and realistic data analysis. 

Our extension is not straightforward given the central role that the latent 
Gaussian field plays in the INLA methodology. In this paper, we take 
advantage of the types of approximations performed and the way they are 
combined in INLA to propose a new way to look at this latent near-Gaussian 
models of interest and show how to adapt INLA to fit this more complex class of models.
The paper is organized as following: Section \ref{sec:INLA} will describe the 
latent Gaussian models and the INLA
methodology, highlighting the importance of the Gaussian latent field to the
success of the method. Section \ref{sec:INLAext} defines the sub-class of
latent models of interest in this paper and describes our proposed extension
to fit this set of models. Section \ref{sec:examples} illustrates our approach
with two examples and Section \ref{sec:concl} offers a conclusion.
The proposed extension is already implemented as part of the R package 
\texttt{INLA}, and its use is illustrated in Appendix \ref{app:survcode},
where the R code from the example in Section \ref{sec:examples:surv}
is displayed. \footnote{Please visit \url{http://www.r-inla.org/} for more information 
about the R package \texttt{INLA}.}  

\section{Integrated Nested Laplace Approximation}\label{sec:INLA}

This section contains a brief description of latent Gaussian models and
a review of the INLA method proposed by \cite{rue2009approximate}.
Latent Gaussian models have a wide range of
applications including, for example, regression models, dynamic models, 
spatial and spatiotemporal models.
In Section \ref{sec:INLAmodels} we define the class of latent Gaussian models 
and its hierarchical representation that will make the exposition of the 
approximation methods described in this paper easier to read. The Gaussian 
approximation to conditional distributions of the latent Gaussian field, which 
is the core of INLA is described in Section \ref{seq:INLA:gaussianapprox}. The 
INLA method applied to latent Gaussian models is described in Section 
\ref{sec:INLAmethod}, while the importance of the Gaussian prior on the latent 
field to the success of INLA is made explicit is Section 
\ref{sec:INLAext-Gaussianmatter}.

\subsection{Latent Gaussian models}\label{sec:INLAmodels}

The INLA framework was designed to deal with latent Gaussian models where the 
observation (or response) variable $y_i$ is assumed to belong to a distribution family 
(not necessarily part of the exponential family)
where some parameter of the family $\phi _i$ is linked to a structured additive 
predictor $\eta _i$ through a link function $g(\cdot)$, so that $g(\phi_i) = \eta_i$.
The structured 
additive predictor $\eta _i$ accounts for effects of various covariates in an 
additive way:
\begin{equation}
\label{eq:strucaddit}
\eta _i = \alpha + \summ{j=1}{n_f} f^{(j)}(u_{ji}) + 
\summ{k=1}{n _{\beta}} \beta _k z_{ki} + \epsilon _i,
\end{equation}
where $\{f^{(j)}(\cdot)\}$'s are unknown functions of the covariates $\bs{u}$, 
used for example to relax linear relationship of covariates and to model temporal
and/or spatial dependence,
the $\{\beta _k\}$'s represent the linear effect of covariates $\bs{z}$ and the 
$\{\epsilon _i\}$'s are unstructured terms. Then a Gaussian prior is assigned 
to $\alpha$, $\{f^{(j)}(\cdot)\}$, $\{\beta _k\}$ and $\{\epsilon _i\}$. 

We can also write the model described above using a hierarchical structure, 
where the first stage is formed by the likelihood function with conditional 
independence properties given the latent field $\bs{x} = (\bs{\eta}, \alpha, 
\bs{f}, \bs{\beta})$ and possible hyperparameters $\bs{\theta}_1$, where each 
data point $\{y_i, i=1,...,n_d\}$ is connected to one element in the latent 
field $x_i$. Assuming that the elements of the latent field connected to the
data points are positioned on the first $n_d$ elements of $\bs{x}$, we have
\begin{list}{\labelitemi}{\leftmargin=5em}
  \item[\textbf{Stage 1.}] $\bs{y}|\bs{x},\bs{\theta}_1 \sim \pi(\bs{y}|\bs{x},\bs{\theta}_1) = 
\prodd{i=1}{n_d} \pi(y_i|x_i,\bs{\theta}_1)$.
\end{list}

Note that the linear predictor $\bs{\eta}$ are usually the first $n_d$ elements of $\bs{x}$,
as represented by the notation $\bs{x} = (\bs{\eta}, \alpha, \bs{f}, \bs{\beta})$ used 
before. The reason to include the linear predictors in the latent field is purely 
computational as it allows the INLA software to be coded for LGMs in general and
not on a case-by-case basis.

The conditional distribution of the latent field $\bs{x}$ given 
some possible hyperparameters $\bs{\theta}_2$ forms the second stage of the 
model and has a joint Gaussian distribution,
\begin{list}{\labelitemi}{\leftmargin=5em}
  \item[\textbf{Stage 2.}] $\bs{x}|\bs{\theta}_2 \sim \pi(\bs{x}|\bs{\theta}_2) = 
\mathcal{N}(\bs{x};\bs{\mu}(\bs{\theta}_2), \bs{Q}^{-1}(\bs{\theta}_2))$,
\end{list}
where $\mathcal{N}(\cdot;\bs{\mu}, \bs{Q}^{-1})$ denotes a multivariate Gaussian 
distribution with mean vector $\bs{\mu}$ and a precision matrix $\bs{Q}$. In most 
applications, the latent Gaussian field have conditional independence properties, 
which translates into a sparse precision matrix $\bs{Q}(\bs{\theta}_2)$, which is 
of extreme importance for the numerical algorithms that will follow. The latent field
$\bs{x}$ may have additional linear constraints of the form $\bs{A}\bs{x} = \bs{e}$ 
for an $k \times n$ matrix $\bs{A}$ of rank $k$, where $k$ is the number of constraints 
and $n$ the size of the latent field. The hierarchical 
model is then completed with an appropriate prior distribution for the hyperparameters 
of the model $\bs{\theta} = (\bs{\theta}_1, \bs{\theta}_2)$
\begin{list}{\labelitemi}{\leftmargin=5em}
  \item[\textbf{Stage 3.}] $\bs{\theta} \sim \pi(\bs{\theta})$.
\end{list}

\subsection{The Gaussian approximation}\label{seq:INLA:gaussianapprox}

The Gaussian approximation to densities of the form
\begin{equation}
 \label{eq:fullcondx}
 \pi(\bs{x}|\bs{\theta}, \bs{y}) \propto \exp\bigg\{ -\frac{1}{2} \bs{x}^T \bs{Q}(\bs{\theta}) \bs{x} 
+ \sum_{i \in I} g_i(x_i) \bigg\},
\end{equation}
plays a important role in INLA, where $g_i(x_i)$ is a function of $x_i$ 
that may depend on $y_i$ and $\bs{\theta}$, and $I$ is an index set. Hence in this section we 
describe one of the many possible ways to approximate (\ref{eq:fullcondx})
by a Gaussian distribution.

We can perform a Taylor expansion up to second order of $g_i(x_i)$
around an initial guess $\mu_i^{(0)}$
\begin{equation*}
 g_i(x_i) \approx g_i(\mu_i^{(0)}) + b_i x_i - \frac{1}{2} c_i x_i^2,
\end{equation*}
where $b_i$ and $c_i$ depend on $\mu_i^{(0)}$, and then a Gaussian approximation is obtained with
precision matrix $\bs{Q} + \text{diag}(\bs{c})$ and mode given by the solution of 
$\{\bs{Q} + \text{diag}(\bs{c})\}\bs{\mu}^{(1)} = \bs{b}$, where $\bs{b}$ and $\bs{c}$ are vectors 
formed by $b_i's$ and $c_i's$ respectively. This process is repeated until it converges
to a Gaussian distribution with, say, mean $\bs{\mu}^*$ and precision matrix 
$\bs{Q} ^* = \bs{Q} + \text{diag}(\bs{c}^*)$, where $\bs{c}^* = \bs{c}(\bs{\mu}^*)$,
which we denote hereafter by $\pi_G(\bs{x}|\bs{\theta}, \bs{y})$.

\subsection{The INLA method}\label{sec:INLAmethod}

For the hierarchical model described in Section \ref{sec:INLAmodels} the 
joint posterior distribution of the unknowns then reads
\begin{align*}
\pi (\bs{x}, \bs{\theta} | \bs{y}) & \propto \pi(\bs{\theta}) \pi(\bs{x}|\bs{\theta}) 
\prodd{i=1}{n_d}\pi(y_i|x_i, \bs{\theta}) \\
& \propto \pi(\bs{\theta})|\bs{Q}(\bs{\theta})|^{n/2} \exp\bigg[ -\frac{1}{2} \bs{x}^T \bs{Q}(\bs{\theta}) \bs{x} + 
\summ{i=1}{n_d} \log \{ \pi(y_i|x_i, \bs{\theta}) \} \bigg]
\end{align*}
The approximated posterior marginals of interest $\tilde{\pi}(x_i|\bs{y})$, $i=1,..,n$ 
and $\tilde{\pi}(\theta _j|\bs{y})$, $j=1,...,m$ returned by INLA has the following form
\begin{align}
\label{eq:INLAximarg}
 \tilde{\pi}(x_i|\bs{y}) & = \sum _k \tilde{\pi}(x_i|\bs{\theta}^{(k)}, \bs{y}) 
 \tilde{\pi}(\bs{\theta}^{(k)}|\bs{y})\ \Delta\bs{\theta} ^{(k)} \\
 \label{eq:INLAthetajmargcont}
 \tilde{\pi} (\theta _j|\bs{y}) & = \int \tilde{\pi} (\bs{\theta}|\bs{y}) d\bs{\theta} _{-j}
\end{align}
where $\{\tilde{\pi}(\bs{\theta} ^{(k)} |\bs{y}) \}$ are the density values computed
during a grid exploration on $\tilde{\pi}(\bs{\theta} |\bs{y})$. Since we do not have 
$\tilde{\pi} (\bs{\theta}|\bs{y})$ evaluated at all
points required to compute the integral in Eq. (\ref{eq:INLAthetajmargcont}) we
construct an interpolation $I(\bs{\theta}|\bs{y})$ using the density values 
$\{\tilde{\pi}(\bs{\theta} ^{(k)} |\bs{y}) \}$ computed
during the grid exploration on $\tilde{\pi}(\bs{\theta} |\bs{y})$ and approximate
(\ref{eq:INLAthetajmargcont}) by
\begin{equation}
 \label{eq:INLAthetajmarg}
 \tilde{\pi}(\theta _j|\bs{y}) = \int I(\bs{\theta}|\bs{y}) d\bs{\theta}_{-j}.
\end{equation}

Looking at [(\ref{eq:INLAximarg})-(\ref{eq:INLAthetajmarg})] we can see that 
INLA can be divided into three main tasks, firstly propose an approximation 
$\tilde{\pi}(\bs{\theta}|\bs{y})$ to the joint
posterior of the hyperparameters $\pi(\bs{\theta}|\bs{y})$, secondly propose an approximation 
$\tilde{\pi}(x_i|\bs{\theta}, \bs{y})$ to the marginals of the conditional distribution of the latent field given the data and 
the hyperparameters $\pi(x_i|\bs{\theta}, \bs{y})$
and thirdly explore $\tilde{\pi}(\bs{\theta}|\bs{y})$ on a grid and use it to integrate out $\bs{\theta}$
in Eq. (\ref{eq:INLAximarg}) and $\bs{\theta}_{-j}$ in Eq. (\ref{eq:INLAthetajmarg}). A simplified 
algorithm would look like the following: 
\begin{enumerate}
\setlength{\itemsep}{-0.5cm}
\setlength{\parskip}{0cm}
\item \vb{Select a set }$\Theta = \{ \bs{\theta}^{(1)}, ..., \bs{\theta}^{(K)} \}$ \\
\item \vb{for }$k = 1, ..., K$\vb{ do} \\
\item \hspace{1cm}\vb{Compute }$\tilde{\pi}(\bs{\theta}^{(k)}|\bs{y})$ \\ 
\item \hspace{1cm}\vb{Compute }$\tilde{\pi}(x_i|\bs{\theta}^{(k)}, \bs{y})$\vb{ as a function of }$x_i$ for all $i$ \\
\item \vb{end for} \\
\item \vb{Compute }$\tilde{\pi}(x_i|\bs{y}) = \sum _k \tilde{\pi}(x_i|\bs{\theta}^{(k)}, \bs{y}) 
 \tilde{\pi}(\bs{\theta}^{(k)}|\bs{y})\ \Delta\bs{\theta} ^{(k)}$\vb{ as a function of }$x_i$\vb{, for all }$i$.
\end{enumerate}
We refer to \cite{rue2009approximate} for details on how to perform the grid exploration 
to obtain $\Theta = \{ \bs{\theta}^{(1)}, ..., \bs{\theta}^{(K)}\}$, since it is not essential for understanding our proposed extension 
described in Section \ref{sec:INLAext}. \cite{martinsbayesian} discuss how to compute 
(\ref{eq:INLAthetajmarg}) efficiently.

The approximation used for the joint posterior of the hyperparameters $\pi(\bs{\theta}|\bs{y})$ is
\begin{equation}
 \label{eq:lapthetay}
 \tilde{\pi} (\bs{\theta}|\bs{y}) \propto 
 \frac{\pi(\bs{x}, \bs{\theta}, \bs{y})}{\pi_G(\bs{x}|\bs{\theta}, \bs{y})}\bigg|_{\bs{x} = \bs{x}*(\bs{\theta})}
\end{equation}
where $\pi_G(\bs{x}|\bs{\theta}, \bs{y})$ is the Gaussian approximation (see Section 
\ref{seq:INLA:gaussianapprox}) to the full conditional of $\bs{x}$, and
$\bs{x}^*(\bs{\theta})$ is the mode of the full conditional for $\bs{x}$, for a given $\bs{\theta}$. 
Expression (\ref{eq:lapthetay}) is equivalent to \cite{tierney1986accurate} Laplace 
approximation of a marginal posterior distribution, and it is exact if $\pi(\bs{x}|\bs{y}, \bs{\theta})$
is a Gaussian.

For $\pi(x_i|\bs{\theta}, \bs{y})$, three options are available, and they vary in 
terms of speed and accuracy. The fastest option, $\pi_G(x_i|\bs{\theta}, \bs{y})$, 
is to use the marginals of the Gaussian approximation $\pi_G(\bs{x}|\bs{\theta}, \bs{y})$
already computed when evaluating expression (\ref{eq:lapthetay}). The only extra cost 
to obtain $\pi_G(x_i|\bs{\theta}, \bs{y})$ is to compute the marginal variances from the sparse
precision matrix of $\pi_G(\bs{x}|\bs{\theta}, \bs{y})$. The Gaussian approximation often 
gives reasonable results, but there can be errors in the location and/or errors due to the 
lack of skewness \citep{rue2007approximate}. The more accurate approach
would be to do again a Laplace approximation, denoted by $\pi_{LA}(x_i|\bs{\theta}, \bs{y})$, 
with a form similar to expression (\ref{eq:lapthetay}) 
\begin{equation}
 \label{eq:lapxidadothetay}
 \pi_{LA}(x_i|\bs{\theta}, \bs{y}) \propto 
\frac{\pi(\bs{x}, \bs{\theta}, \bs{y})}{\pi_{GG}(\bs{x}_{-i}|x_i, \bs{\theta}, \bs{y})}
\bigg|_{\bs{x}_{-i} = \bs{x}_{-i}*(x_i,\bs{\theta})},
\end{equation}
where $\pi_{GG}(\bs{x}_i|x_i, \bs{\theta}, \bs{y})$ is the Gaussian approximation to 
$\bs{x}_i|x_i, \bs{\theta}, \bs{y}$ and $\bs{x}_{-i}*(x_i,\bs{\theta})$ is the modal 
configuration. A third option $\pi_{SLA}(x_i|\bs{\theta}, \bs{y})$, called simplified 
Laplace approximation, is obtained by doing a Taylor expansion on the numerator and 
denominator of expression (\ref{eq:lapxidadothetay}) up to third order, thus correcting 
the Gaussian approximation for location and skewness with a much lower cost when compared to 
$\pi_{LA}(x_i|\bs{\theta}, \bs{y})$. We refer to \cite{rue2009approximate} for a 
detailed description of the Gaussian, Laplace and simplified Laplace approximations to
$\pi(x_i|\bs{\theta}, \bs{y})$.

\subsection{INLA and the importance of the Gaussian field}\label{sec:INLAext-Gaussianmatter}

The main challenge in applying INLA to latent models is that the method depends heavily on
the latent Gaussian prior assumption to work properly, both from the computational point of view and
from the choice of approximations used as described in Section \ref{sec:INLAmethod}.

For the full conditional of $\bs{x}$ to be well approximated by a Gaussian distribution in
equations (\ref{eq:lapthetay}) and (\ref{eq:lapxidadothetay}), we need
it to be well behaved and close to a Gaussian. This is basically ensured by the latent Gaussian prior that is assigned 
to $\bs{x}$ (see Stage 2 of Section \ref{sec:INLAmodels}) in latent Gaussian models, which has a non-negligible effect on the posterior, especially
in terms of dependence between the components of $\bs{x}$. 

Another important issue is that the conditional independence properties
often encountered in the latent field translates into a sparse precision matrix when it is Gaussian distributed. This
implies a huge decrease in computational time when performing the Gaussian approximation, which is extremely important
since a Gaussian approximation needs to be computed for each value $\bs{\theta}^{(k)}$ used on the grid for the numerical
integration in Eq. (\ref{eq:INLAximarg}).

\section{Extension to near-Gaussian latent models}\label{sec:INLAext}

In this section we show how to extend the scope of INLA to include models similar in structure
to the latent Gaussian models described in Section \ref{sec:INLAmodels} but where the prior
for some components of the latent field can have a near-Gaussian distribution. 
Section \ref{sec:INLAext:models} will define these latent models in
general while Section \ref{sec:INLAext:1stex} will present an specific example, namely 
a survival model with gamma frailty. The proposed extension is then presented in Section
\ref{sec:INLAext:main}.

\subsection{Near-Gaussian latent models}\label{sec:INLAext:models}

The models we are interested in this paper has the same structure as the latent Gaussian models described
in Section \ref{sec:INLAmodels} with the exception that the latent field has some independent non-Gaussian
components. We redefine stage 2 of the hierarchical model of Section \ref{sec:INLAmodels} as
\begin{list}{\labelitemi}{\leftmargin=5em}
  \item[\textbf{Stage $\bs{2^{new}}$.}] $\underbrace{(\bs{x}_{G},\bs{x}_{NG})}_{\bs{x}}|\bs{\theta}_2 \sim \pi(\bs{x}|\bs{\theta}_2) = \mathcal{N}(\bs{x}_G;\bs{0}, \bs{Q}^{-1}(\bs{\theta}_2)) \times \prod _i \pi(\bs{x}_{NGi}|\bs{\theta}_2)$,
\end{list}
where $\bs{x}_{G}$ and $\bs{x}_{NG}$ represent the Gaussian and non-Gaussian terms of the latent field, respectively.
In addition we assume that $\bs{x}_{NG}$ is formed by independent random variables. As a result,
the distribution of the latent field is not Gaussian anymore, which precludes the use of INLA to
fit this class of models. 


The term \emph{near-Gaussian} latent models refer to the
restrictions we impose on the non-Gaussian components of the latent
field. We aim for non-Gaussian distributions that are not too different
from a Gaussian one, and are so that they could be well enough
approximated by a Gaussian density. 
Although it is hard to give a precise definition of which distributions 
belong to the class of near-Gaussian distributions, we can restrict the 
possibilities by considering only distributions in which the density have 
full support on the real line, a unique and unimodal mode, finite first
two moments and decreasing density as we move away from the mode.
The main point here is to understand that our main interest in this paper lies on distributions 
that add flexibility around a Gaussian distribution, by correcting it in terms 
of skewness and/or kurtosis.

There are two main reasons for these restrictions. Firstly, it will imply that 
our proposed methodology returns accurate approximations for the posterior 
marginals of the non-Gaussian components, as will be shown in Section 
\ref{sec:INLAext:main}.
Secondly, it fits
the framework described in Section \ref{sec:intro} in which we state that our 
ultimate goal is to include options in the \texttt{R-INLA} package so that
the user could expand an initial model $M_0$ based on Gaussian distributions 
in certain directions by adding flexibility around those Gaussian distributions, 
e.g. correcting it for asymmetry and/or kurtosis. This framework
is an ongoing project 
and involves 
other considerations besides the computational methods presented here.
For example, in this setting, we want the prior distributions for the 
hyperparameters to be chosen so that the extended model $M_1$ could effectively 
be interpreted as an extension of the basic model $M_0$, i.e. in a way that
$M_0$ would have a central role within $M_1$. 
The INLA extension described in Section \ref{sec:INLAext:main} is an essential
component of this framework of more flexible models within \vb{R-INLA}, and since
the near-Gaussian distribution concept has a central role on this framework, it 
becomes important to understand that the concern here is with the design of
algorithms that works well on this context of near-Gaussian latent models.

There is no clear way to diagnose if a given non-Gaussian distribution 
fits the class of near-Gaussian distributions for the purpose of obtaining accurate
results with our extension. Our experience have been that it works well for the cases 
we are currently interested in, which are distributions that correct the Gaussian in
terms of skewness and/or kurtosis. However, the success of our extension has been 
verified on a case-by-case basis. This is not different than what have been done for
most of the deterministic approximate methods for Bayesian inference. Our suggestion is to perform
simulation studies to verify the accuracy of our approximations for each new class
of non-Gaussian distribution that one might be interested.

\subsection{Survival model - A first example}\label{sec:INLAext:1stex}

Consider the following exponential model that can be used to analyze
survival data that comes from subjects of the same group who are related to each other 
or from multiple recurrence times of a event for the same individual. The likelihood 
\begin{equation}
 \label{eq:ex:multisurv:likelihood}
 t_{ij} \sim \exp(\lambda _{ij}), \quad i=1,...,I\mbox{ and }j=1,...,J
\end{equation}
is given by independent exponential distributions given $\bs{\lambda} = \{\lambda _{ij},\ i=1,...,I,\ j=1,...,J\}$,
where $\lambda _{ij}=1/\mu_{ij}$ and $\mu_{ij}$ is the mean of $t_{ij}$. The index $I$ could be interpreted
as the number of groups in the data, while $J$ would be the number of individuals in each group. This is a case
of balanced data-set, but the unbalanced case could be treated just as easily by our method. 

It is expected that individuals belonging to the same group are correlated with each other. This can
be included in the model through the addition of random effects $\bs{w} = \{w_1, ..., w_I\}$ to account
for variation between groups, 
\begin{equation}
 \label{eq:ex:multisurv:linearpred}
 \eta_{ij} = \log(\lambda _{ij}) = \beta _0 + \beta _1 x_{ij} + \log w_i.
\end{equation}
Besides the random effects, it is common to include some covariate effects that in our case are represented by the
fixed effects $\beta_0$ and $\beta_1$. In the survival analysis literature, the random effects are often called
frailty and it is common to assume that they are Gamma distributed, $w_i \sim \text{Gamma}(\kappa, \kappa)$, 
with $E(w_i) = 1$ to avoid identifiability issues. 
Notice that we have $\log w_i$ in Eq. 
(\ref{eq:ex:multisurv:linearpred}) only because this is how the model is defined in practice. That is,
$\lambda_{ij} = \exp\{\beta_0 + \beta_1\}w_i$, hence Eq. (\ref{eq:ex:multisurv:linearpred}). This is not to be
confused with the application of log transformation to make the random effects' distribution closer
to a Gaussian.
Gaussian priors are assumed for the fixed effects and a Gamma prior is often used for the 
random-effect hyperparameter $\kappa$.

\begin{figure}[ht!]
  \centering
  \includegraphics[width=0.6\linewidth]{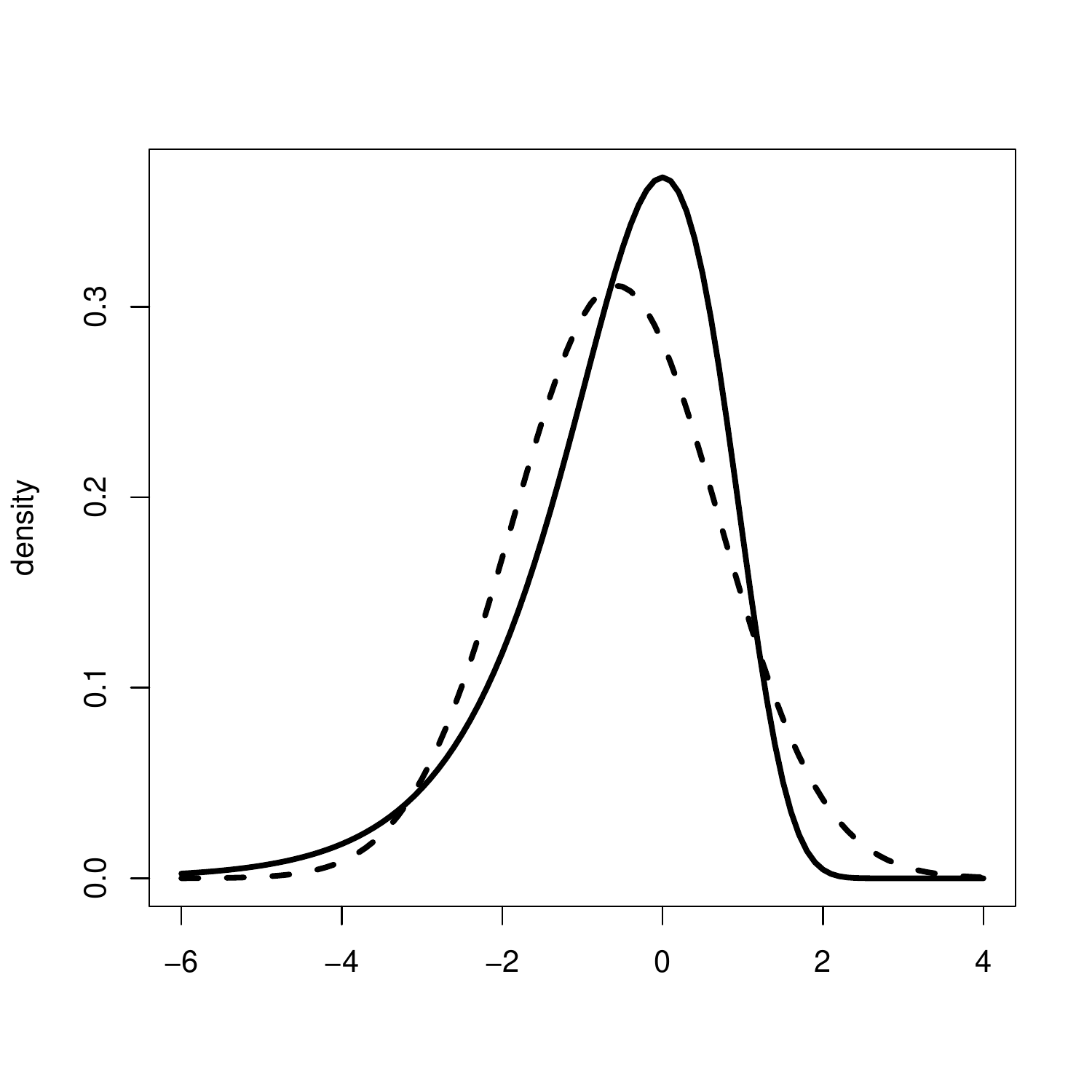}
  \caption{Log-gamma density (solid) with $\kappa = 1$ and a Gaussian density (dashed)
  with the same mean and precision as the log-gamma density.}
\label{fig:log_gamma_density}
\end{figure}

The latent field for this model is given by $\bs{x} = (\bs{\eta}, \bs{\beta}, \bs{b})$, with 
$\bs{b} = \log (\bs{w})$ and it is non Gaussian since $\bs{b}$ is formed by independent 
log-Gamma random variables,
\begin{equation}
 \label{eq:ex:multisurv:randomeffect}
 \pi(\bs{b}|\kappa) = \prodd{i=1}{I} \pi(b_i|\kappa) = \prodd{i=1}{I} \frac{\kappa ^\kappa}{\Gamma(\kappa)} \exp \{ \kappa (b_i - \exp(b_i)) \}. 
\end{equation}
Such a model cannot be applied straightforwardly using INLA since the assumption in Stage 2 is violated. 
However, it fits the class of models in Section \ref{sec:INLAext:models} and will be further analyzed in Section \ref{sec:examples:surv} with our approach. 
Figure \ref{fig:log_gamma_density} shows a log-gamma distribution with $\kappa = 1$
(solid line) and a Gaussian density (dashed line) with the same mean and precision as the
log-gamma with $\kappa = 1$. We can see in Figure \ref{fig:log_gamma_density} that this
log-gamma has negative skewness, positive kurtosis, and satisfy the desired properties 
of a near-Gaussian distribution as described in Section \ref{sec:INLAext:models}.

\subsection{The extension}\label{sec:INLAext:main}

Given that INLA does not require the prior for $\bs{\theta}$ to be Gaussian, 
a first approach to fit the class of models presented in Section \ref{sec:INLAext:models} 
with INLA would be to include the non-Gaussian components 
$\bs{x}_{NG}$ in the hyperparameters
$\bs{\theta}$ of the model. However, this is not a good idea in practice since the size of
$\bs{x}_{NG}$ is usually large and typically increase as the number of data points. This
naive approach would lead to accurate results but the cost would be a large increase in
computational time due to the grid exploration necessary to compute Eqs. (\ref{eq:INLAximarg})-(\ref{eq:INLAthetajmarg}). 
Basically, an increase in the dimension of $\bs{\theta}$ would require a higher number
of points $\bs{\Theta} = \{\bs{\theta}^{(1)}, ..., \bs{\theta}^{(K)}\}$ to be evaluated 
during the grid exploration of $\bs{\theta}$.
The approach presented next will deliver accurate results without the burden in 
computational time.

\subsubsection{The main idea}

Section \ref{sec:INLAext-Gaussianmatter} explained the importance
of the latent Gaussian prior for INLA to run smoothly. With that in mind we propose to approximate the 
prior of the non-Gaussian components $\pi(\bs{x}_{NG}|\bs{\theta}_2)$ by a Gaussian distribution 
$\pi_G(\bs{x}_{NG}|\bs{\theta}_2)$ and correct this approximation by the correction term 
\begin{equation}
 \label{eq:ext:CT}
 CT = \pi(\bs{x}_{NG}|\bs{\theta}_2)/\pi_G(\bs{x}_{NG}|\bs{\theta}_2) 
\end{equation}
in the likelihood. 
This is, in fact, a way of writing a latent model of the form described in Section \ref{sec:INLAext:models}
into a latent Gaussian model, defined in Section \ref{sec:INLAmodels}. 

The first stage is now formed by the original likelihood multiplied by the correction term
\begin{equation*}
\prodd{i=1}{n_d} \pi(y_i|x_i,\bs{\theta}_1) \times \pi(\bs{x}_{NG}|\bs{\theta}_2)/\pi_G(\bs{x}_{NG}|\bs{\theta}_2).
\end{equation*}
Another way of writing this is to define an extended response vector $\bs{z}$, with $z_i = y_i$ if
$i \leq n_d$ and $z_i = 0$ if $n_d < i \leq n_d + k$, where $k$ is the length of $\bs{x}_{NG}$ and write
\begin{list}{\labelitemi}{\leftmargin=5em}
  \item[\textbf{Stage 1.}] $\bs{z}|\bs{x},\bs{\theta} \sim \pi(\bs{z}|\bs{x}, \bs{\theta}) = \prodd{i=1}{n_d + k} \pi(z_i|x_i, \bs{\theta})$, where
\end{list}
\begin{equation}
\label{eq:fakelikelihood}
 \pi(z_i|x_i, \bs{\theta}) = \Bigg\{ \begin{array}{cc}
                                      \pi(y_i|x_i,\bs{\theta}_1) & \mbox{ for }1 \leq i \leq n_d \\
                                      \pi(x_{NG,i}|\bs{\theta}_2)/\pi_G(x_{NG,i}|\bs{\theta}_2) & \mbox{ for } n_d < i \leq n_d+k
                                     \end{array}
\end{equation}

It is important to emphasize that Stage 1 above is not the likelihood function, but expressing the model
using this form will make the description and implementation of the algorithm that follows easier.
The latent field has now a Gaussian approximation replacing the non-Gaussian distribution of $\bs{x}_{NG}$,
\begin{list}{\labelitemi}{\leftmargin=5em}
  \item[\textbf{Stage 2.}] $\underbrace{(\bs{x}_{G},\bs{x}_{NG})}_{\bs{x}}|\bs{\theta}_2 \sim \pi(\bs{x}|\bs{\theta}_2) = 
\mathcal{N}(\bs{x}_G;\bs{\mu}(\bs{\theta}_2), \bs{Q}^{-1}(\bs{\theta}_2)) \times \pi_G(\bs{x}_{NG}|\bs{\theta}_2)$,
\end{list}
which means that now $\pi(\bs{x}|\bs{\theta}_2)$ is Gaussian distributed. The third stage is once again formed
by the prior distribution on the hyperparameters,
\begin{list}{\labelitemi}{\leftmargin=5em}
  \item[\textbf{Stage 3.}] $\bs{\theta} \sim \pi(\bs{\theta})$.
\end{list}

Independent of the Gaussian approximation $\pi_G(\bs{x}_{NG}|\bs{\theta}_2)$ used, the hierarchical model
above is equivalent to the model described in Section \ref{sec:INLAext:models}. Considerations on
how to choose this Gaussian approximation and how this model
formulation will help us to perform inference will be presented soon, but first we can rewrite the survival model of Section \ref{sec:INLAext:1stex} in this new formulation.

\subsubsection*{Survival model (Cont.)}

For the survival model defined in Section \ref{sec:INLAext:1stex},
we have that the original likelihood function, defined in Eq. (\ref{eq:ex:multisurv:likelihood}), 
is an exponential distribution
\begin{equation*}
 \log \pi(t_{ij}|\eta_{ij}) = \eta_{ij} - \exp(\eta_{ij} t_{ij}),
\end{equation*}
where $\eta_{ij}$ is the linear predictor defined in Eq. (\ref{eq:ex:multisurv:linearpred}).
Based on Eq. (\ref{eq:ex:multisurv:randomeffect}) the correction term (see Eq. (\ref{eq:ext:CT})) is given by
\begin{align*}
CT & = \prodd{i=n_d + 1}{n_d + I}\pi(\bs{x}_{NG,i}|\bs{\theta}_2)/\pi_G(\bs{x}_{NG,i}|\bs{\theta}_2) \\
& = \prodd{i=1}{I} \pi(b_i|\kappa)/\pi_G(b_i|\mu_b, \tau_b) = \prodd{i=1}{I} CT_i,
\end{align*}
with
\begin{equation}
 \label{eq:ex:multisurv:logCTi}
 \log CT_i = \kappa(b_i - \exp(b_i)) + \frac{\tau _b(\kappa)}{2} (b_i - \mu _b(\kappa))^2 + \text{const},
\end{equation}
where $\mu_b(\kappa)$ and $\tau_b(\kappa)$ are the mean and precision parameter of the Gaussian
approximation to the log-Gamma random effects $\bs{b}$ and const is a constant that does not depend on $\bs{b}$. 
The latent field 
$\bs{x} = (\bs{\eta}, \bs{b}, \bs{\beta})$ is now Gaussian since $\pi(\bs{b}|\kappa)$
is approximated by $\pi_G(\bs{b}|\mu_b(\kappa), \tau_b(\kappa)) = \mathcal{N}(\bs{b};\mu_b(\kappa)\bs{1}_I, \tau_b(\kappa)^{-1}\bs{I}_I)$,
where $\bs{1}_n$ is a vector of ones with dimension $n$ and $\bs{I}_n$ is an $n\times n$ identity matrix.

\subsubsection{Computational considerations}\label{sec:comp_considerations}

At first sight, it seems obvious that once our latent (non-Gaussian) model of interest
has been rewritten into a latent Gaussian model we could apply INLA to obtain the
posterior marginals of interest. However, we need to understand what are the 
consequences of this change within the INLA framework.
The main change is on 
the Gaussian approximation to the full conditional of the latent field (see Section 
\ref{seq:INLA:gaussianapprox}), that now takes the form
\begin{equation}
 \label{eq:INLAext:fullcondx}
 \pi(\bs{x}|\bs{\theta}, \bs{y}) \propto \exp\bigg\{ -\frac{1}{2} \bs{x}^T \bs{Q}(\bs{\theta}) \bs{x} 
+ \summ{i=1}{n_d} g_i(x_i) + \summ{i = n_d+1}{n_d+k} h_i(x_i) \bigg\},
\end{equation}
where $g_i(x_i) = \log \pi(y_i|x_i,\bs{\theta})$ as before and
\begin{equation*}
h_i(x_i) = \log CT_i = \log \pi(\bs{x}_{NG,i}|\bs{\theta}_2) - \log \pi_G(\bs{x}_{NG,i}|\bs{\theta}_2).
\end{equation*}
It was shown in Section \ref{seq:INLA:gaussianapprox} that a Gaussian approximation is obtained
by approximating the non-quadratic functions, in this case $g_i(x_i)$ and $h_i(x_i)$, by quadratic 
functions using Taylor expansion up to 
second order. Once we know we are dealing with a well behaved log likelihood
function $g_i(x_i)$ as, for example, those belonging to the exponential family, the success of
a Gaussian approximation to Eq. (\ref{eq:INLAext:fullcondx}) will depend heavily on the shape
of $h_i(x_i)$. 

For instance, it is desirable to have a bounded correction term 
\begin{equation*}
 \pi_{NG}(\cdot|\bs{\theta})/\pi_G(\cdot|\bs{\theta}) < \infty
\end{equation*}
for a quadratic form approximation to $h_i(x_i)$ to make sense. This implies that the Gaussian
approximation $\pi_G(\bs{x}_{NG}|\bs{\theta})$ should ideally dominate 
$\pi_{NG}(\bs{x}_{NG}|\bs{\theta})$ in the sense that it should have thicker tails 
than the non-Gaussian distribution it is trying to approximate. But in practice, it is
sufficient to have a bounded correction term on the region that concentrates the bulk of probability
mass since we can afford a bigger approximation error on the region that doesn't contribute
much to the density (\ref{eq:INLAext:fullcondx}). In our examples, we have chosen 
$\pi_G(\cdot|\bs{\theta})$ to be a Gaussian distribution with zero mean ($\mu = 0$)
and low precision ($\tau \rightarrow 0$) to approximate the distribution of the 
non-Gaussian components $\pi_{NG}(\cdot|\bs{\theta})$. This choice satisfy the bounded
correction term requirement and since it does not depend on $\bs{\theta}$, it avoids the complication of computing a new Gaussian approximation for each value of
$\bs{\theta}$ in the grid exploration.

Although not necessary, it is desirable to have both the correction term $\exp \{h_i(x_i)\}$
and the likelihood function $\exp\{g_i(x_i)\}$ to be log-concave, at least on a neighborhood
of the mode of Eq. (\ref{eq:INLAext:fullcondx}). This would imply $\pi(\bs{x}|\bs{\theta}, \bs{y})$ 
defined in Eq. (\ref{eq:INLAext:fullcondx}) to be log-concave. It is easier to design an 
algorithm to maximize a concave function. For example, a quadratic model function
provides a good local approximation to the objective function being maximized.
This can be seen in Section \ref{seq:INLA:gaussianapprox} where we propose to maximize 
Eq. (\ref{eq:fullcondx}) by performing a series of quadratic approximations. The mode
$\bs{\mu}^*$ is then found iteratively by solving the linear system
\begin{equation}
 \label{eq:NRcommon}
 \bs{Q}'(\bs{\mu}^{(j-1)}) \bs{\mu}^{(j)} = \bs{b}(\bs{\mu}^{(j-1)}) 
\end{equation}
until convergence is attained, where $\bs{Q}'(\bs{\mu}^{(j-1)}) = \{ \bs{Q} + \text{diag}(\bs{c}(\bs{\mu}^{(j-1)}))\}$
and $\bs{Q}$, $\bs{b}$ and $\bs{c}$ are defined in Section 
\ref{seq:INLA:gaussianapprox}. This is equivalent to a Newton-Raphson algorithm.

Next, we show that the survival model in Section \ref{sec:INLAext:1stex} is one example
where both the correction term and the likelihood function are log-concave.

\subsubsection*{Survival model (Cont.)}

\begin{figure}[ht!]
  \centering
    \subfigure[]
     {\label{fig:multisurv-loglik}\includegraphics[width=0.35\linewidth]{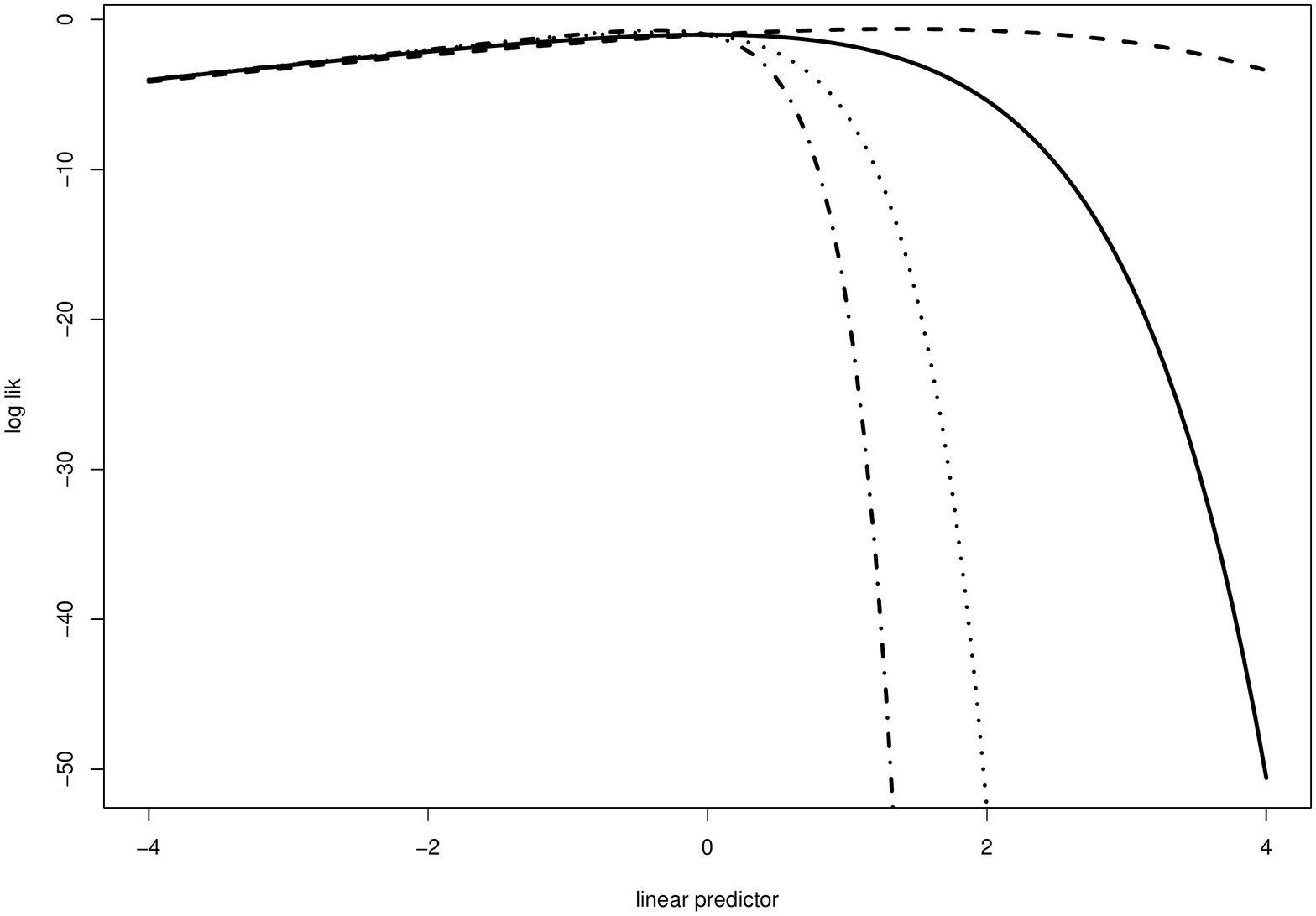}}
    \subfigure[]
     {\label{fig:multisurv-deriv2loglik}\includegraphics[width=0.35\linewidth]{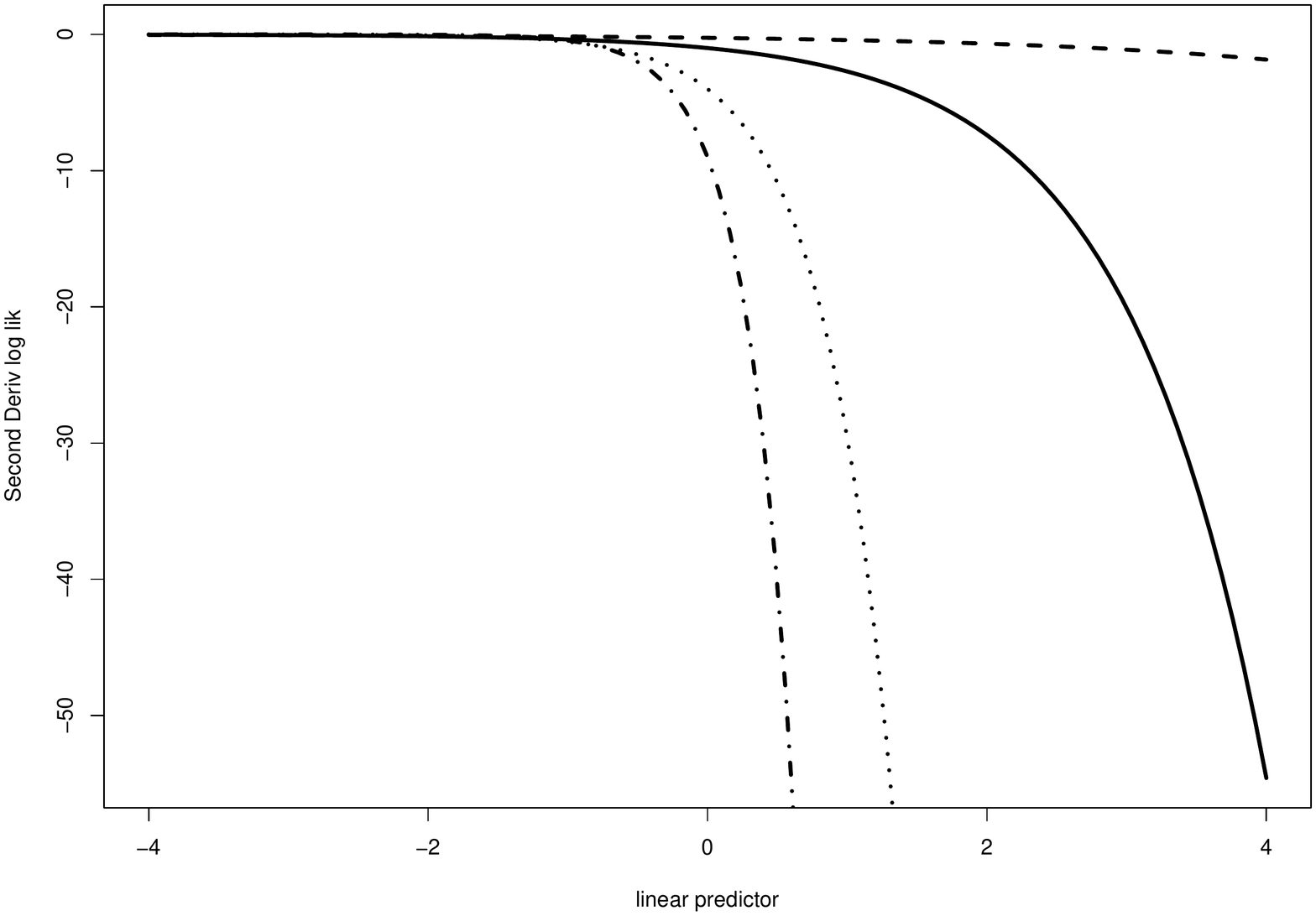}}
  \caption{Plot of the log likelihood (a) and of the second derivative of the log likelihood (b) against 
the linear predictor for different values of data points for the survival model. It was used the following 
values for the data point: 0.5 (Dashed), 1 (Solid), 2 (Dotted) and 3 (Dot-dash)}
\label{fig:multisurv-likelihood}
\end{figure}

For the survival model, we have a log-concave likelihood function as we can see
in Figure \ref{fig:multisurv-likelihood}, where we have the plots of the log likelihood and 
the second derivative of the log likelihood for a given data point, assuming different
values for the data point.

\begin{figure}[ht!]
  \centering
    \subfigure[]
     {\label{fig:multisurv-loglik}\includegraphics[width=0.35\linewidth]{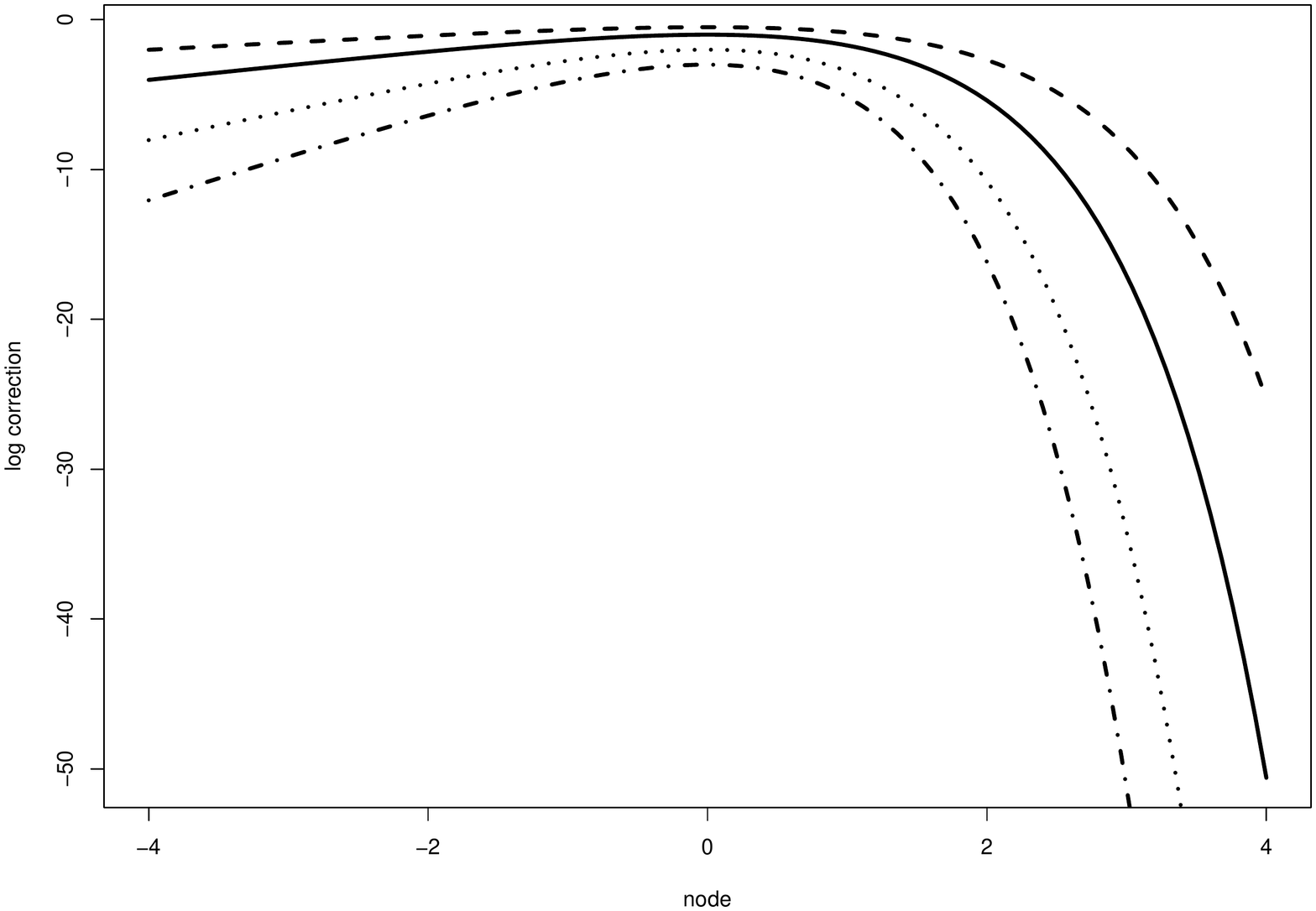}}
    \subfigure[]
     {\label{fig:multisurv-deriv2loglik}\includegraphics[width=0.35\linewidth]{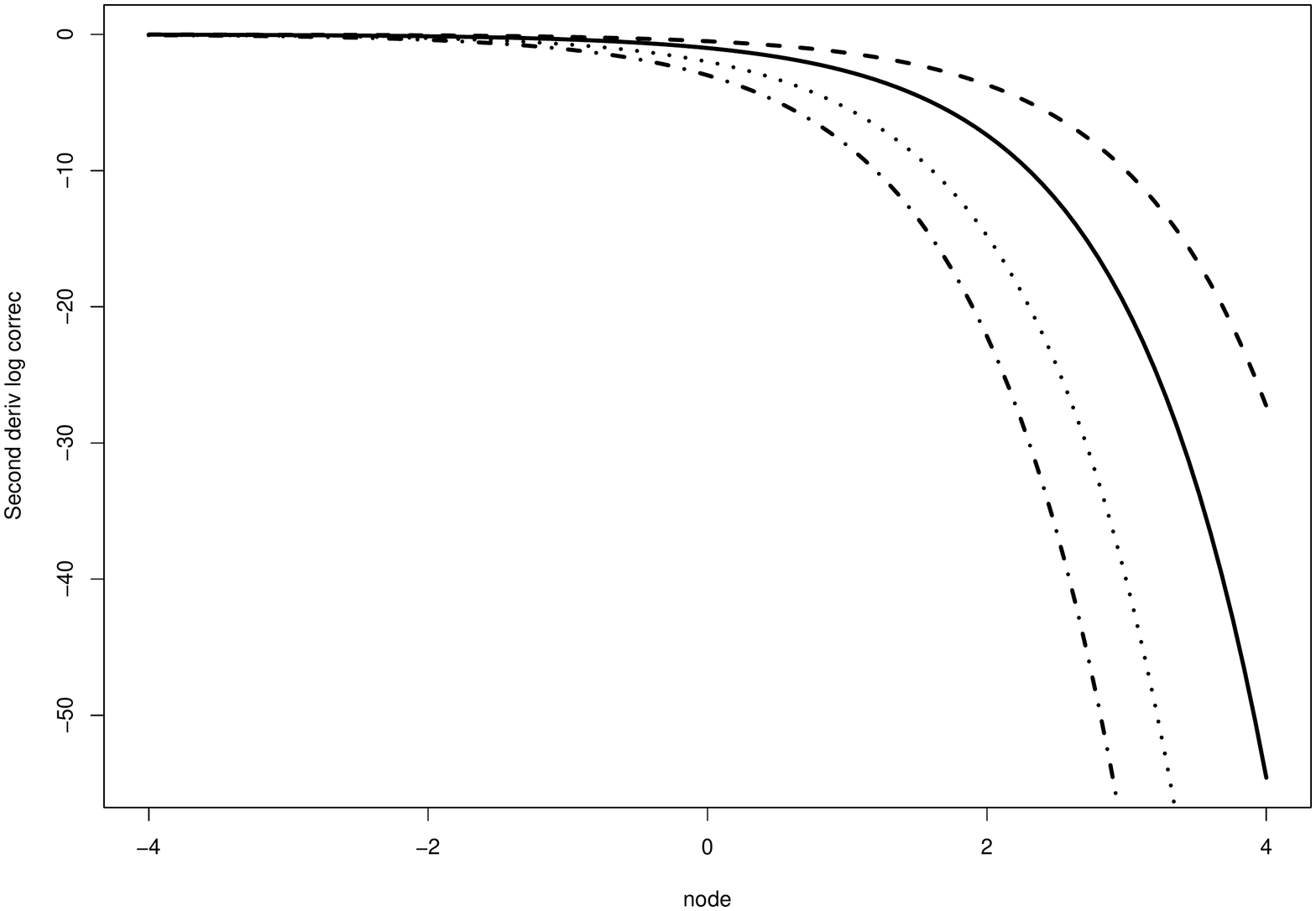}}
  \caption{Plot of the log correction term (a) and of the second derivative of the log correction term (b) 
against $b_i$ for different values of $\kappa$ for the survival model. It was used the following values 
for $\kappa$: 0.5 (Dashed), 1 (Solid), 2 (Dotted) and 3 (Dot-dash)}
\label{fig:multisurv-correcterm}
\end{figure}

If we use a zero mean and low precision Gaussian distribution ($\mu_b = 0$ and $\tau _b \rightarrow 0$) in 
Eq. (\ref{eq:ex:multisurv:logCTi}) we also attain a log-concave correction term, 
\begin{equation}
 \label{eq:ex:multisurv:logCTisimpl}
 \log CT_i = \kappa(b_i - \exp(b_i)) + \text{const}
\end{equation}
as illustrated in Figure \ref{fig:multisurv-correcterm}.

\begin{flushright}
$\square$
\end{flushright}

The example in Section \ref{sec:examples:robustMM} considers a linear mixed-effects
model where both the likelihood and the random effects are distributed according to
a Student's t distribution. This is a case where both the likelihood and the correction
term are not log-concave. This brings an extra challenge for the optimization of
$\pi(\bs{x}|\bs{\theta},\bs{y})$ in Eq. (\ref{eq:INLAext:fullcondx}). We have
obtained good results using trust region methods \citep{conn1987trust}, which are
well known in the optimization literature. In our context, we constrain or expand the
subset of the domain, known as trust region, where a quadratic function is used to
approximate the objective function $\pi(\bs{x}|\bs{\theta},\bs{y})$ to be optimized.

We accomplish this by iteratively solving 
\begin{equation*}
 \{\bs{Q}'(\bs{\mu}^{(j-1)}) + \delta \text{diag}(\bs{Q}'(\bs{\mu}^{(j-1)}))\} \bs{\mu}^{(j)} = \bs{b}(\bs{\mu}^{(j-1)}) 
\end{equation*}
instead of Eq. (\ref{eq:NRcommon}), where $\delta$ is the parameter that control the
trust-region size. At each iteration we expand the trust region by decreasing $\delta$ 
if the quadratic approximation to $\pi(\bs{x}|\bs{\theta},\bs{y})$ is adequate. On the
other hand, we shrink the trust region by increasing $\delta$ if the quadratic approximation
is not adequate. When we get close to the mode $\bs{\mu}^*$, the quadratic function 
becomes a good approximation to $\pi(\bs{x}|\bs{\theta},\bs{y})$, in which case we can
set $\delta$ to zero and return to the original optimization problem given by Eq. 
(\ref{eq:NRcommon}). 

Trust-region methods have a nice Bayesian interpretation in this context. The use of $\delta > 0$ in this case
could be interpreted as an increase in the precision of the prior of the latent field. This solution is in agreement with the findings of \cite{vanhatalo2009gaussian}. They have found, in the context of a Student's t likelihood, that the most problematic case in the optimization of a non log-concave full conditional of the type (\ref{eq:INLAext:fullcondx}) happens when the prior is much wider than the likelihood, in which case moderate prior-data conflict can lead to numerical instability. By increasing the precision of the prior in the moments of numerical instability, we allow the algorithm to proceed until a point close to the mode, where $\delta$ is set to zero and the original prior is recovered.

\subsubsection{Checking accuracy}\label{sec:check_accuracy}

We can use the same tools described in \cite{rue2009approximate} to assess
the approximation error of our approach. The accuracy of $\tilde{\pi}(\bs{\theta}|\bs{y})$
is directly related to the ``effective" dimension of the latent field $\bs{x}$.
The package \vb{R-INLA} returns the expected number of effective parameters, $\text{Eeff}$.
In general, if $\text{Eeff} < n_d/2$ we have strong evidence that the Gaussian approximation
to $\pi(\bs{x}|\bs{\theta}, \bs{y})$ in Eq. (\ref{eq:INLAext:fullcondx}) is adequate,
where $n_d$ is the number of data points.
 
The second strategy is based on the idea of comparing elements of a sequence 
of increasingly accurate approximations. By default, \vb{R-INLA} computes the
symmetric Kullback-Leibler divergence (SKLD) between the integrated marginals
in Eq. (\ref{eq:INLAximarg}) obtained using the Gaussian and the simplified 
Laplace approximation to $\pi(x_i|\bs{\theta}, \bs{y})$, respectively. If
the divergence is small then both approximations are considered as acceptable.
Otherwise, one should compute Eq. (\ref{eq:INLAximarg}) using the Laplace
approximation to $\pi(x_i|\bs{\theta}, \bs{y})$ and check the SKLD between
this and the one obtained with the simplified Laplace approximation.
Again, if the divergence is small, simplified Laplace and Laplace approximations 
appear to be acceptable; otherwise, the Laplace approximation is our best estimate,
but further investigation might be required to check the quality of the
approximations.

Needless to say, those strategies are not fail proof. The only
way to assess with certainty the approximation of our approach would 
be to run a MCMC sampler for an infinite time; and even if this was 
possible, we would need a way to check with certainty if the MCMC chain have
converged to the correct stationary distribution, which is also
not an easy task.

\section{Examples}\label{sec:examples}

We now proceed to two examples where we apply the methodology proposed in this paper and 
compare the results with that obtained by MCMC. Comparison with MCMC has no practical
value and is presented here only to convince the reader that with only a small fraction 
of computational time our approach gives similar accuracy when compared with MCMC algorithms. 
In practice, we suggest to use the tools presented in Section \ref{sec:check_accuracy} 
to check the accuracy of the approximations. The INLA software is in constant
development, but support regarding the code used in this paper can be found 
at the INLA website (\url{http://www.r-inla.org/}).

\subsection{Survival analysis with Gamma frailty}\label{sec:examples:surv}

Here we apply our proposed extension to fit the model defined in Section \ref{sec:INLAext:1stex}
in a simulated data-set and compare the results with that obtained by MCMC. For the experiment
reported we have simulated $100$ groups, each of which with $10$ individuals. The covariates
$\{x_{ij}\}$ in Eq. (\ref{eq:ex:multisurv:linearpred}) were simulated from a uniform distribution on the
interval $(0,1)$ while the frailties came from a Gamma distribution with both the shape and rate 
parameters equal to $1$. The fixed effects $\beta_0$ and $\beta_1$ were chosen to be $1$.

\begin{figure}[ht!]
  \centering
    \subfigure[]
     {\label{fig:survmeanlogfrailty}\includegraphics[width=0.45\linewidth]{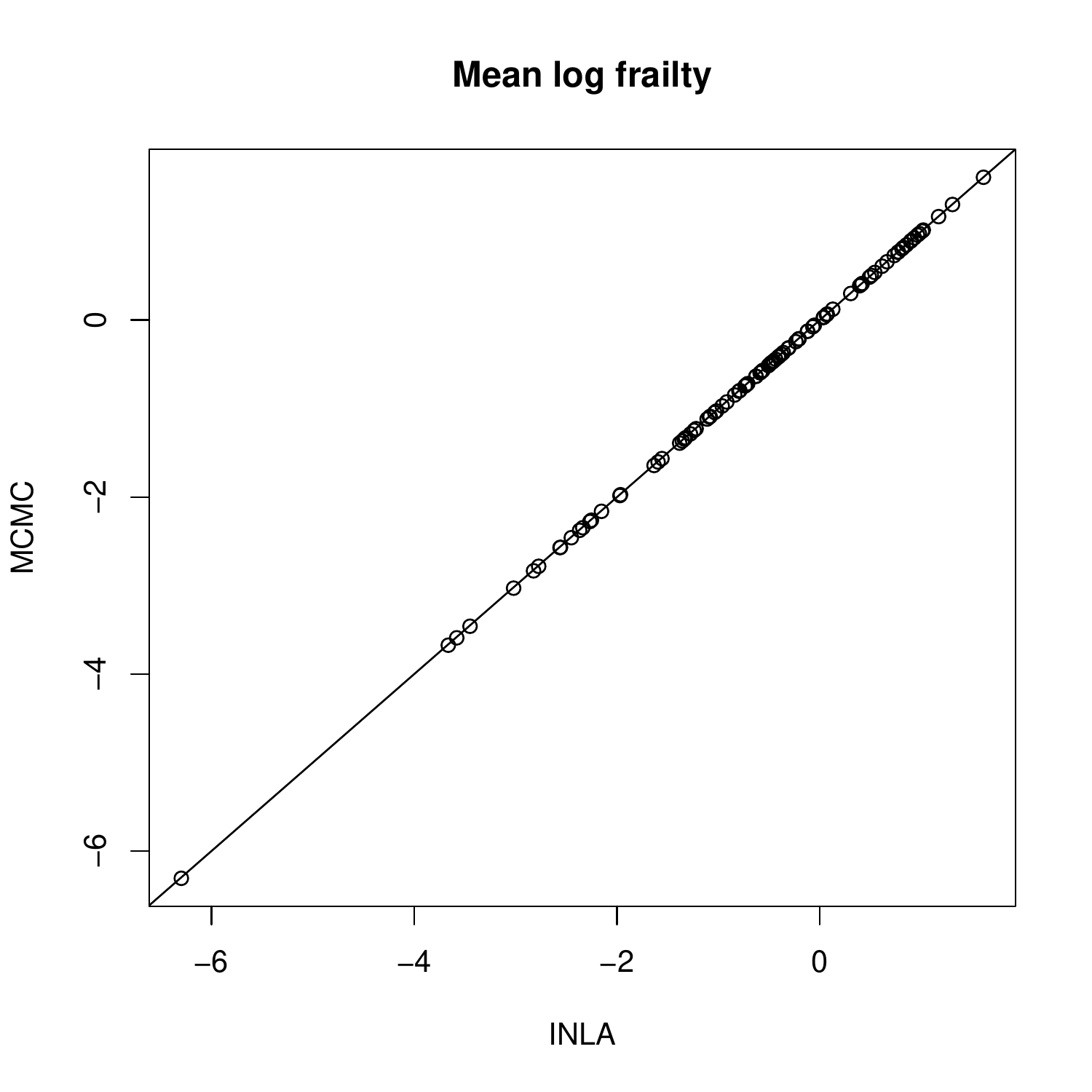}}
    \subfigure[]
     {\label{fig:survlogw80}\includegraphics[width=0.45\linewidth]{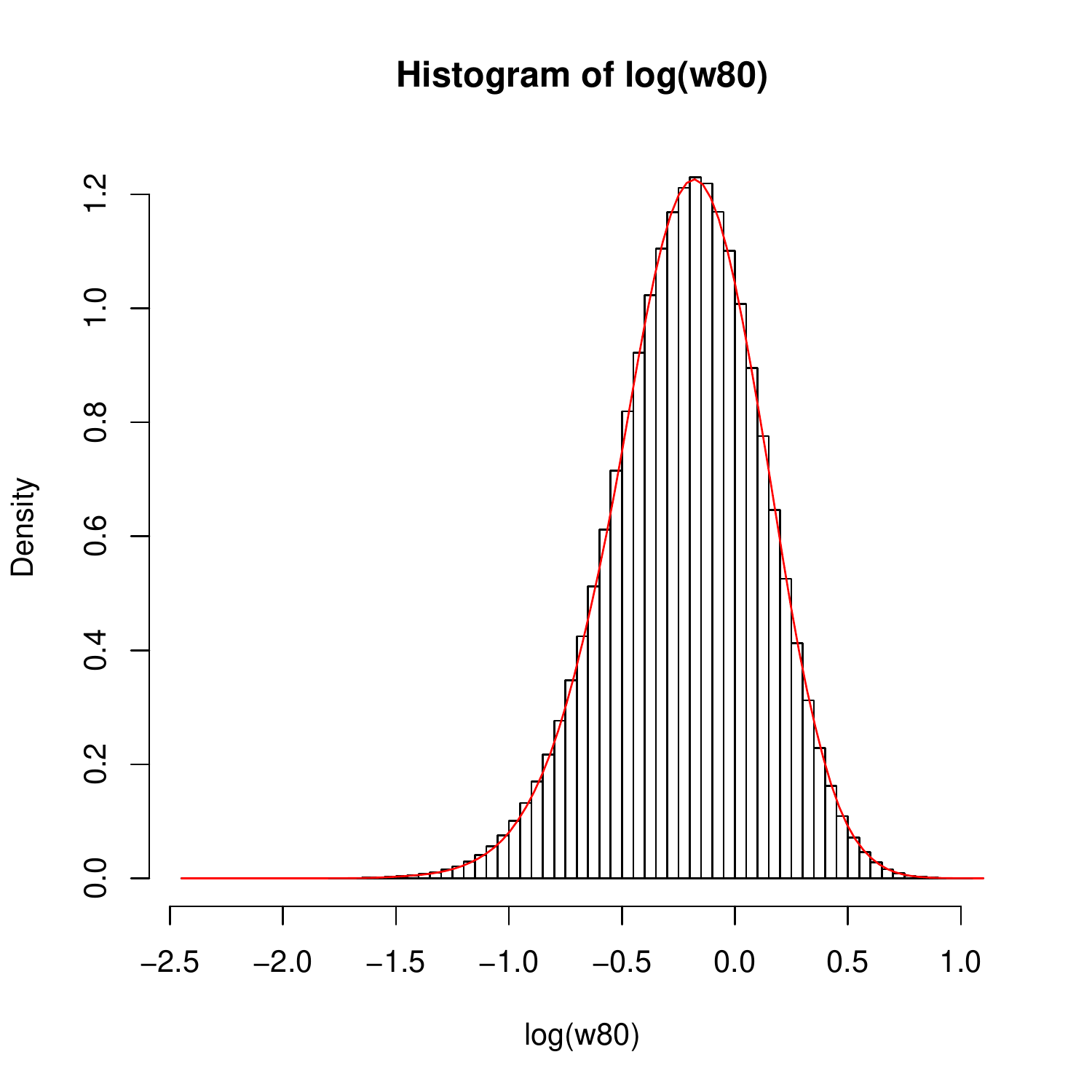}}
  \caption{Comparison between INLA and MCMC for the exponential gamma frailty example: 
(a) Plot of the posterior mean of the log frailties returned by INLA (x-axis) vs. MCMC (y-axis).
(b) Approximate posterior density for $\log w_{80}$ obtained by INLA (solid line) and by MCMC (histogram).  
}
\label{fig:surv1}
\end{figure}

\begin{figure}[ht!]
  \centering
    \subfigure[]
     {\includegraphics[width=0.45\linewidth]{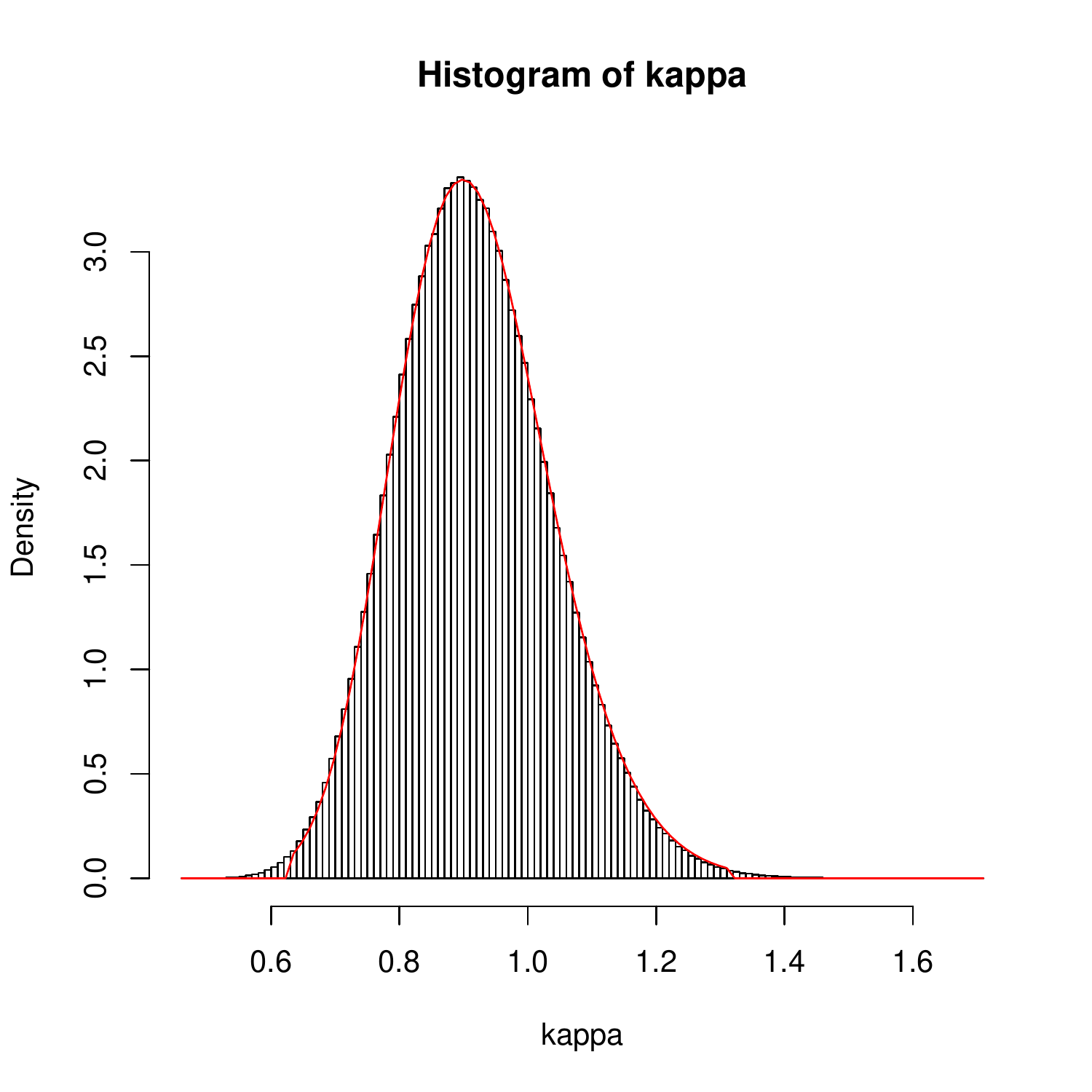}\label{fig:survkappa}}
    \subfigure[]
     {\includegraphics[width=0.45\linewidth]{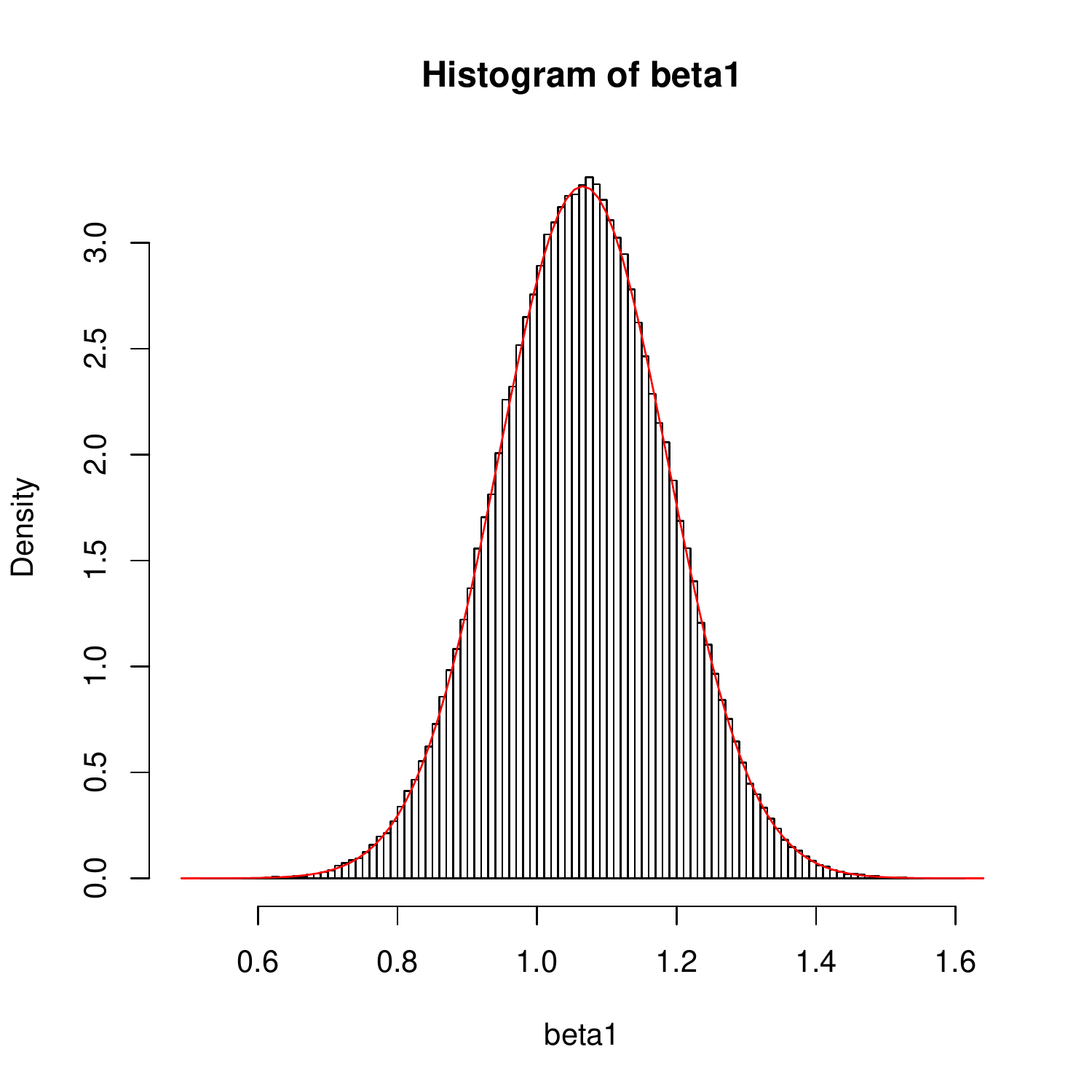}\label{fig:survbeta1}}
  \caption{Comparison between INLA and MCMC for the exponential gamma frailty example:
(a) Approximate posterior density for $\kappa$ obtained by INLA (solid line) and by MCMC (histogram)
(b) Approximate posterior density for $\beta _1$ obtained by INLA (solid line) and by MCMC (histogram)}
\label{fig:surv2}
\end{figure}

We use OpenBUGS \citep{lunn2009bugs} to generate samples from the posterior
distribution. Figure \ref{fig:survmeanlogfrailty} shows the posterior mean of the 
log frailties $\{b_i : b_i = \log w_i,\ i=1,...,100\}$ obtained by INLA (x-axis) and
by MCMC (y-axis). An identity function is also plotted in order to help visualize 
the strong agreement between both methods. Figure \ref{fig:survlogw80} display the 
histogram formed by the samples of $\pi(b_{80}|\bs{y})$ returned by OpenBUGS and the line
is the approximated posterior computed using our extension. This specific component was chosen at
random, since similar accuracy was obtained for all log frailties in our simulation study.
Figure \ref{fig:surv2} show similar pictures for $\beta_1$ and $\kappa$ to show that the excellent
results are also valid for the fixed effects and for the hyperparameter $\kappa$ of our model.
The R code used to run INLA in this example is available in Appendix \ref{app:survcode}.

It is important to note that the comparisons with MCMC were made with
millions of samples, taking minutes to run, since for short to medium number of samples it was
possible to visually detect errors in the MCMC estimates when compared to INLA, that took a
little bit more than 1 second to run on a Intel Core i5 with 2.67GHz. One can argue that the
number of samples (and time) necessary by a MCMC scheme to attain the desired accuracy of 
our application could be reduced if more time was spent designing a specific MCMC scheme 
for this particular application instead of using the general purpose OpenBUGS. Although
this is true in theory, we are here comparing two general purpose tools for the class of 
latent models into consideration, and even with a tailored MCMC scheme, we believe that
the difference in time will still be in orders of magnitude, not to mention the time
necessary to develop specific algorithms for each new model belonging to this same class.

\subsection{Robust mixed-effects models using Student-t distribution}\label{sec:examples:robustMM}

In this example, we show our method applied to a Bayesian random effects model where
the random effects have an Student-t distribution.
Assume we have data $\{\bs{y}_i;\ i=1,...,n\}$ recorded for $n$ groups each having $k_i$ individuals. 
Assume that $\bs{y}_i's$ are independent Gaussian random vector described by the following standard mixed effect model 
\citep{laird1982random}, useful to analyze repeated measures or grouped data,
\begin{equation}
 \label{eq:rmm:model1}
 \bs{y}_i = \bs{X}_i\bs{\beta} + b_i + \bs{e}_i,
\end{equation}
where both the random effect $b_i$ and the error term $\bs{e}_i$ have a Gaussian distribution, $b_i \sim N(0,\sigma_b^2)$ 
and $e_i \sim N(0,\sigma_e^2\bs{I})$ with variances $\sigma^2_b$ and $\sigma_e ^2$ respectively, being $\bs{I}$ a 
($k_i \times k_i$) identity matrix. $X_i$ represent the 
$(k_i \times p)$ design matrix for group $i$ and $\beta$ is a $(p \times 1)$ vector of fixed effects. 

Statistical inference based on the Gaussian distribution is known to be vulnerable to outliers. One
approach to more robust modeling is to replace the Gaussian distribution by Student-t distribution in
the model. In the context of linear mixed effects model, \cite{pinheiro2001efficient} suggested to follow the
robust statistical modeling approach described by \cite{lange1989robust} in which the Gaussian
distributions of $b_i$ and $e_i$ are replaced by $t$-distributions,
\begin{equation}
 \label{eq:rmm:model2}
 b_i \sim t(0, \psi^2_b, \nu), \quad \bs{e}_i \sim t(0, \psi^2_e\bs{I}, \nu),
\end{equation}
where $\psi^2_b$ and $\psi^2_e$ are the scale parameters and $\nu$ is the common degree of freedom parameter.
They also noted that in mixed effects models the outlier may occur either at the level of within-group
error $\bs{e}_i$, called $\bs{e}$-outliers, or at the level of random effects $b_i$, called $\bs{b}$-outliers.
This approach can be regarded as outlier accommodation although it provides useful information for 
outlier identification. For the simulation experiment performed later in this Section, we have used
a Gamma prior with shape and rate parameters given by $1$ and $0.1$ respectively for the inverse scale parameters, 
a Gaussian distribution with mean zero and low precision ($10^{-4}$) for the fixed effects and a Gaussian distribution
with mean $3$ and variance $1$ for $\nu^* = \log (\nu - 5)$, so that the bulk of prior probability mass is between
7 and 150 for the degree of freedom parameter $\nu$. Note that we have defined $\nu ^*$ so that $\nu > 5$ in order to get
a well defined first four moments of the Student-t distribution.

\begin{figure}[ht!]
  \centering
    \subfigure[]
     {\label{fig:tmm-deriv2loglik}\includegraphics[width=0.35\linewidth]{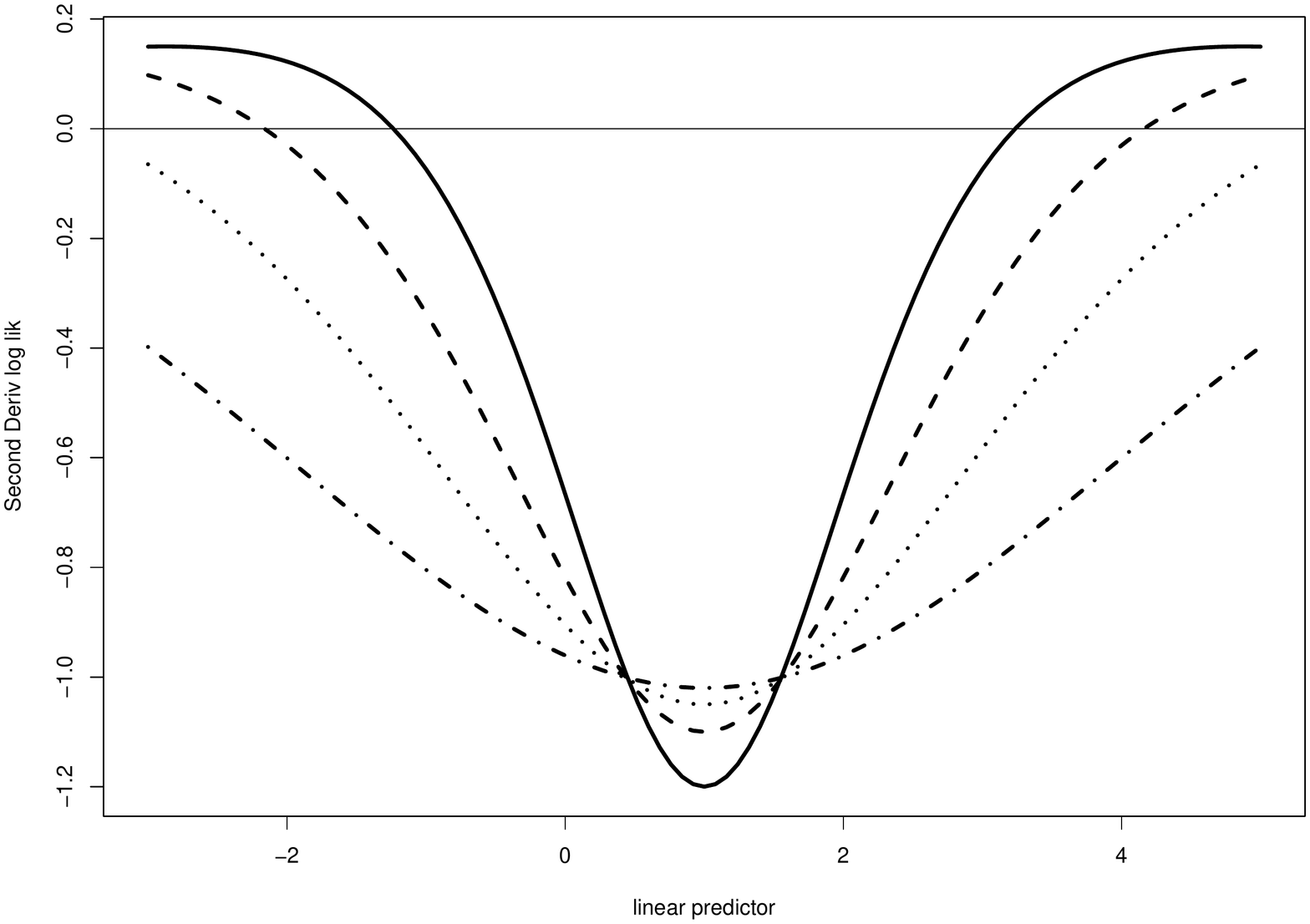}}
    \subfigure[]
     {\label{fig:tmm-deriv2correcterm}\includegraphics[width=0.35\linewidth]{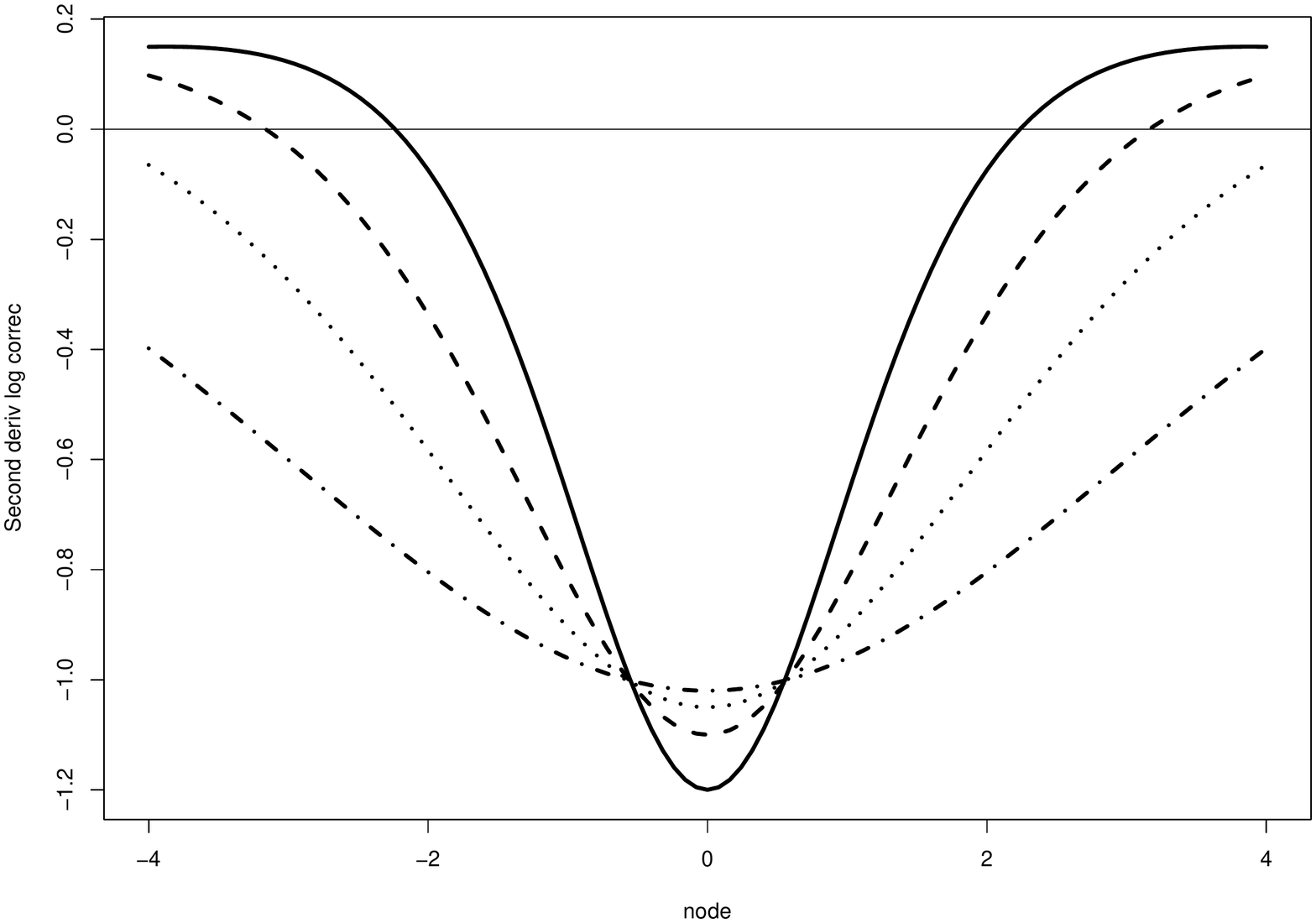}}
  \caption{(a) Plot of the second derivative of the log likelihood against linear predictor for $\psi^2_e$ = 1, $y=1$ and different values of $\nu$ and 
(b) plot of the second derivative of the log correction term against $b_i$ for $\psi_b^2 = 1$ and different values of $\nu$ for the t- mixed effect model.
It was used the following values for $\nu$: 5 (Solid), 10 (Dashed), 20 (Dotted) and 50 (Dot-dash).}
\label{fig:tmm-correcterm}
\end{figure}

The model (\ref{eq:rmm:model1})-(\ref{eq:rmm:model2}) has the likelihood function formed by a $t$ distribution
\begin{equation*}
 y_{ij}|\bs{x}, \bs{\theta} \sim t(\eta_{ij}, \psi ^2_e, \nu), \quad i = 1,...,n \mbox{ and } j=1,...,k_i, 
\end{equation*}
which does not belong to the exponential family, where $\eta_{ij}$ is the linear predictor
\begin{equation*}
 \eta_{ij} = \bs{X}_i\bs{\beta} + b_i.
\end{equation*}
The latent field is then formed by $\bs{x} = (\bs{\eta}, \bs{b}, \bs{\beta})$, where
$\bs{b} = \{b_i;\ i=1,...,n\}$ is formed by independent $t$ distributed random variables and 
has therefore a non-Gaussian distribution given by
\begin{equation}
\label{eq:rmm:randomdens}
\pi(\bs{b}|\psi_b^2, \nu) = \prod_{i=1}^{n} \pi(b_i|\psi_b^2, \nu) = \prod_{i=1}^{n} \frac{\Gamma(\frac{\nu + 1}{2})}{\Gamma(\frac{\nu}{2})} (\psi_b^2 \pi \nu)^{-1/2} \bigg[1 + \frac{b_i^2}{\psi_b^2 \nu}\bigg]^{-(\nu+1)/2}
\end{equation}
If we use Eq. (\ref{eq:ext:CT}) and (\ref{eq:rmm:randomdens}) we get the following log correction
term
\begin{align*}
 \log CT_i & = \log \pi(b_i|\psi_b^2, \nu) - \log \pi_G(b_i|\mu_b, \tau_b) \\
& = -\frac{(\nu+1)}{2} \log \bigg\{ 1+\frac{b_i^2}{\psi_b^2 \nu} \bigg\} + \frac{\tau _b}{2} (b_i - \mu _b)^2 + \text{const}.
\end{align*}
Again, if we assume a zero mean and low precision Gaussian distribution ($\mu_b = 0$ and $\tau_b \rightarrow 0$) we end up
with
\begin{align*}
 \log CT_i = -\frac{(\nu+1)}{2} \log \bigg\{ 1+\frac{b_i^2}{\psi_b^2 \nu} \bigg\} + \text{const}.
\end{align*}
Figure \ref{fig:tmm-correcterm}
show plots of the second derivative of the likelihood (Figure \ref{fig:tmm-deriv2loglik}) and 
of the log correction (Figure \ref{fig:tmm-deriv2correcterm})
term assuming a data point $y = 1$, variances $\psi^2_e = 1$, $\psi_b^2 = 1$ and different values for $\nu$ 
($\nu = 5, 10, 20, 50$).

We have here an example where both the likelihood and the correction term are not log-concave, specially for
low values of degree of freedom parameter $\nu$. 
As explained in Section \ref{sec:comp_considerations}, our extension can be applied 
to models where the likelihood and/or correction term are not log-concave. We use
trust region methods to make the optimization of $\tilde{\pi}(\bs{x}|\bs{\theta}, \bs{y})$
more robust.

Now we can proceed to a contamination study similar to that performed in \cite{pinheiro2001efficient}.
Here we have data simulated from
\begin{equation}
y_i = X\beta + b_i + e_i, \quad i = 1, ..., 27, \quad X = \left[\begin{array}{cc} 1 & 8 \\ 1 & 10 \\ 1 & 12 \\ 1 & 14 \end{array}\right]
\end{equation}
with the following mixture of Gaussian models being used to contaminate the distributions of
the $b_i$ and the $e_i$.
\begin{align*}
b_i & \overset{ind}{\sim} (1-p_b) \mathcal{N}(0, \sigma_b^2) + p_b f \mathcal{N}(0, \sigma_b^2) \\
e_i & \overset{ind}{\sim} (1-p_e) \mathcal{N}(0, \sigma_e^2) + p_e f \mathcal{N}(0, \sigma_e^2), \quad i = 1,...,27, \quad j=1,...,4 
\end{align*}
where $p_b$ and $p_e$ denote, respectively, the expected percentage of $\bs{b}$- and $\bs{e}$-outliers in
the data and $f$ denotes the contamination factor. The true parameters for the uncontaminated 
distributions are $\sigma_b ^2 = 3$ and $\sigma_e ^2 = 2$, while the true values for the fixed
effects are $\beta = (12, 1)^T$.

All $32$ combinations of $p_b, p_e = 0, .05, .1, .25$, and $f =2,4$ were used in the simulation study.
The $f = 2$ case corresponds to a close contamination pattern, while $f = 4$ illustrates
a more distant contamination pattern. A total of $500$ Monte Carlo replications were obtained
for each $(p_b,\ p_e,\ f)$ combination.

Let $\theta$ denote a parameter of interest, with target value $\theta_0 \neq 0$, estimated by 
$\hat{\theta}$, which in our case will be the posterior mean of $\theta$. The efficiency of
the Gaussian estimator $\hat{\theta}_G$ relative to the multivariate $t$ estimator $\hat{\theta}_T$
is defined as the ratio of the respective mean square errors,
\begin{equation}
\label{eq:tmm-effic}
E(\hat{\theta}_G - \theta _0)^2 / E(\hat{\theta}_T - \theta _0)^2,  
\end{equation}
where expectations are
taken with respect to the simulation distribution, that is 
$\widehat{E(\hat{\theta} - \theta _0)^2} = \sum _{i=1}^{500}(\hat{\theta}_i - \theta _0)^2/500$.

\begin{figure}[ht!]
  \centering
    \subfigure[Plot of the posterior mean of the log random effects returned by INLA (x-axis) vs. MCMC (y-axis).]
     {\label{fig:tmm-logb}\includegraphics[width=0.45\linewidth]{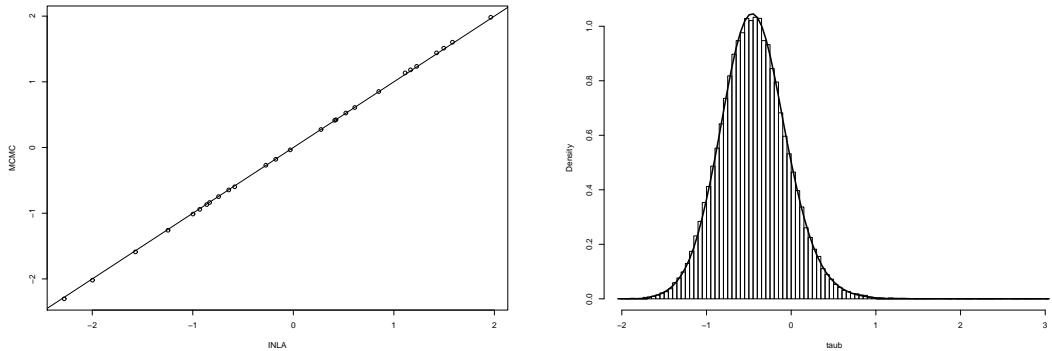}}
    \subfigure[{Approximate posterior density for $\log \tau_b = \log 1/\psi_b^2$ obtained by INLA (solid line) and by MCMC (histogram)}]
     {\includegraphics[width=0.45\linewidth]{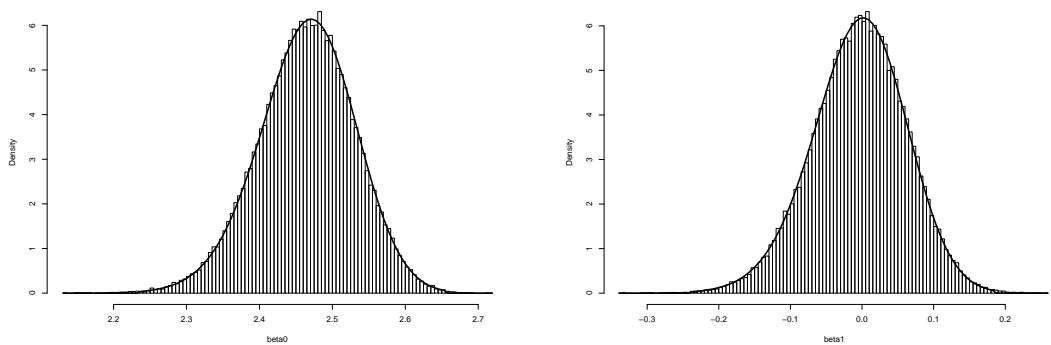}\label{fig:tmm-taub}}
  \caption{Comparison between INLA and MCMC for the robust mixed effect model.}
\label{fig:tmm1}
\end{figure}

\begin{figure}[ht!]
  \centering
    \subfigure[{Approximate posterior density for $\log \beta_0$ obtained by INLA (solid line) and by MCMC (histogram)}]
     {\label{fig:tmm-beta0}\includegraphics[width=0.45\linewidth]{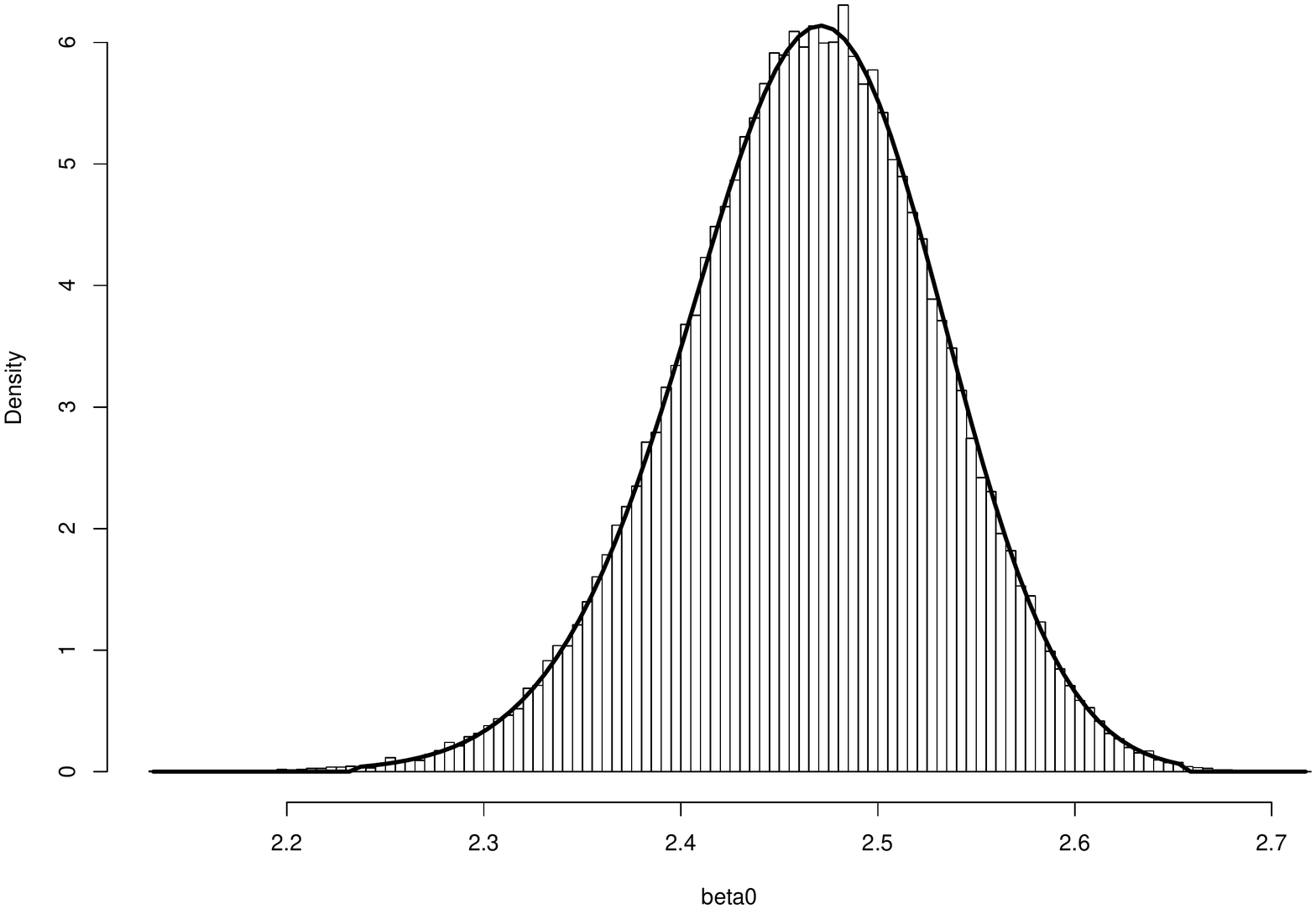}}
    \subfigure[{Approximate posterior density for $\log \beta_1$ obtained by INLA (solid line) and by MCMC (histogram)}]
     {\label{fig:tmm-beta1}\includegraphics[width=0.45\linewidth]{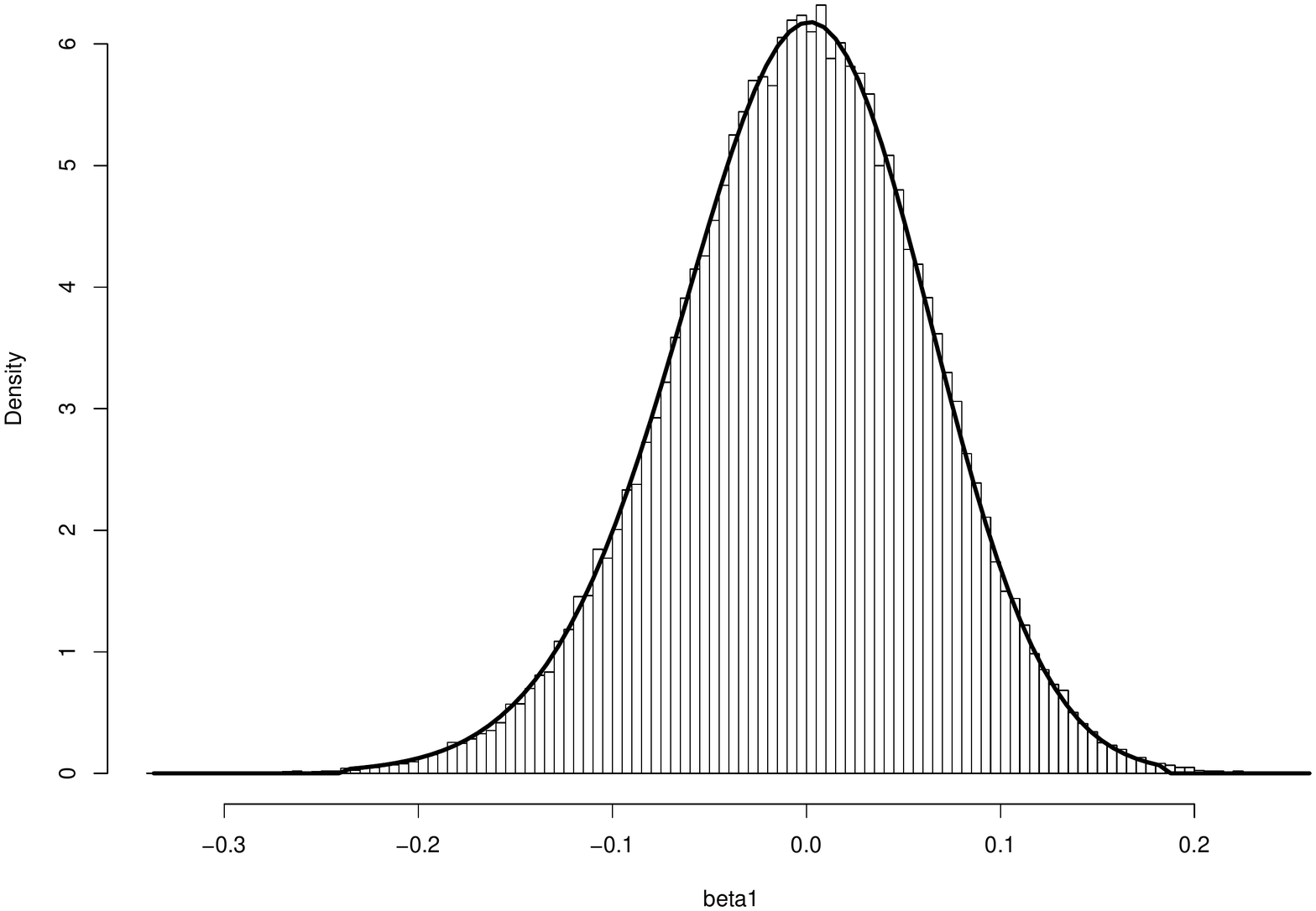}}
  \caption{Comparison between INLA and MCMC for the robust mixed effect model.}
\label{fig:tmm2}
\end{figure}

We have chosen some data-sets out of the $32 \times 500 = 16000$ used in this contamination 
study and fitted the model using both MCMC and INLA to make sure that INLA is doing at
least as good as MCMC in terms of accuracy. After that we proceed with the contamination
study with INLA as the only estimation method as it would be impractical to fit all $16000$
data-sets with MCMC. Figures \ref{fig:tmm1} and \ref{fig:tmm2} illustrate this
comparison for one of the data-sets, where Figure \ref{fig:tmm-logb} display the log random 
effects returned by INLA (x-axis) vs. MCMC (y-axis), while Figures \ref{fig:tmm-taub}
and \ref{fig:tmm2} display the approximate posterior densities for $\log \tau_b = \log 1/\psi_b^2$ 
and for the log fixed effects $\log \beta_0$ and $\log \beta_1$ respectively, obtained by INLA (solid line) 
and by MCMC (histogram). 

\begin{figure}[ht!]
  \centering
    \subfigure[$\beta_0$ under close contamination pattern ($f = 2$)]
     {\label{fig:eficbeta0f2}\includegraphics[width=0.45\linewidth]{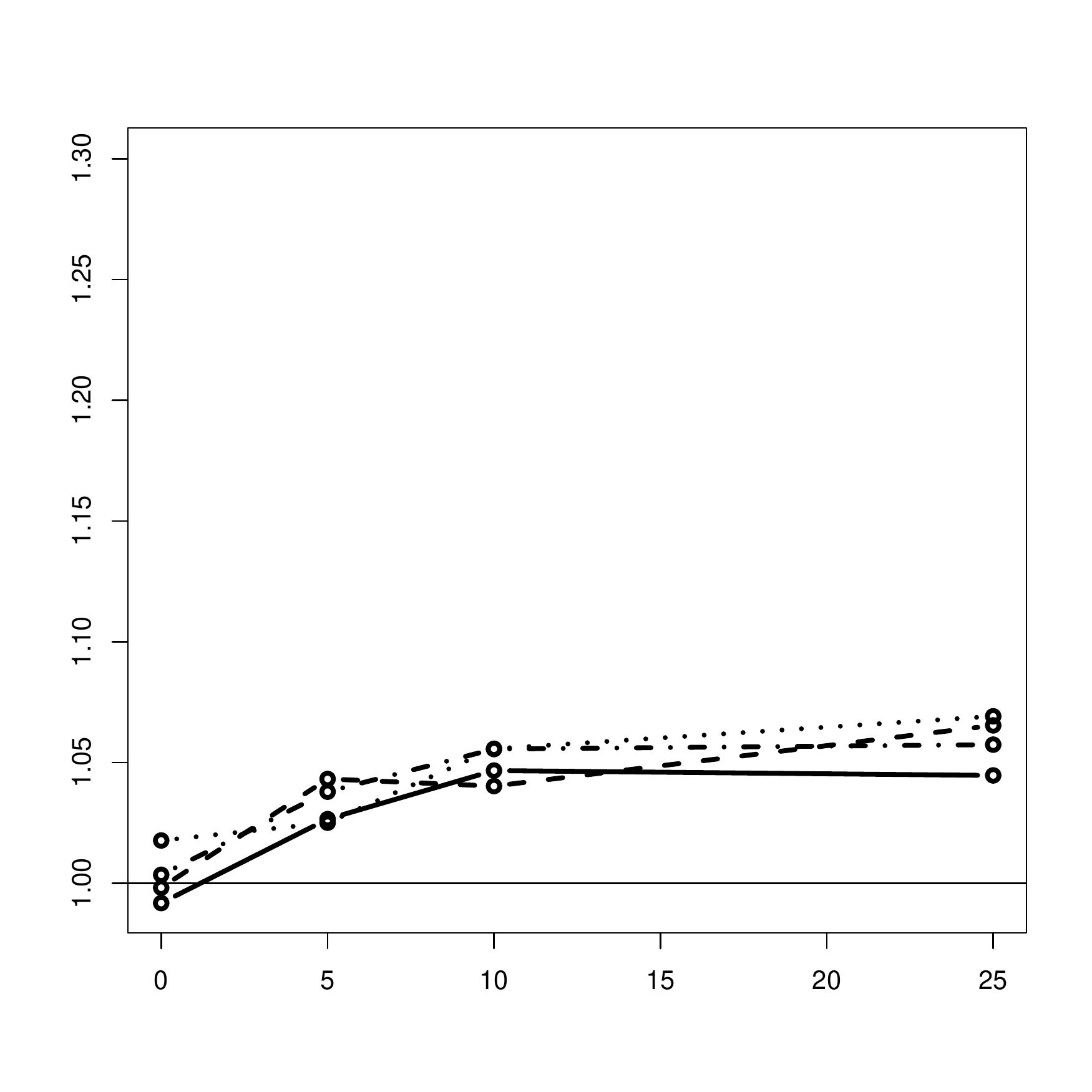}}
    \subfigure[$\beta_0$ under distant contamination pattern ($f = 4$)]
     {\label{fig:eficbeta0f4}\includegraphics[width=0.45\linewidth]{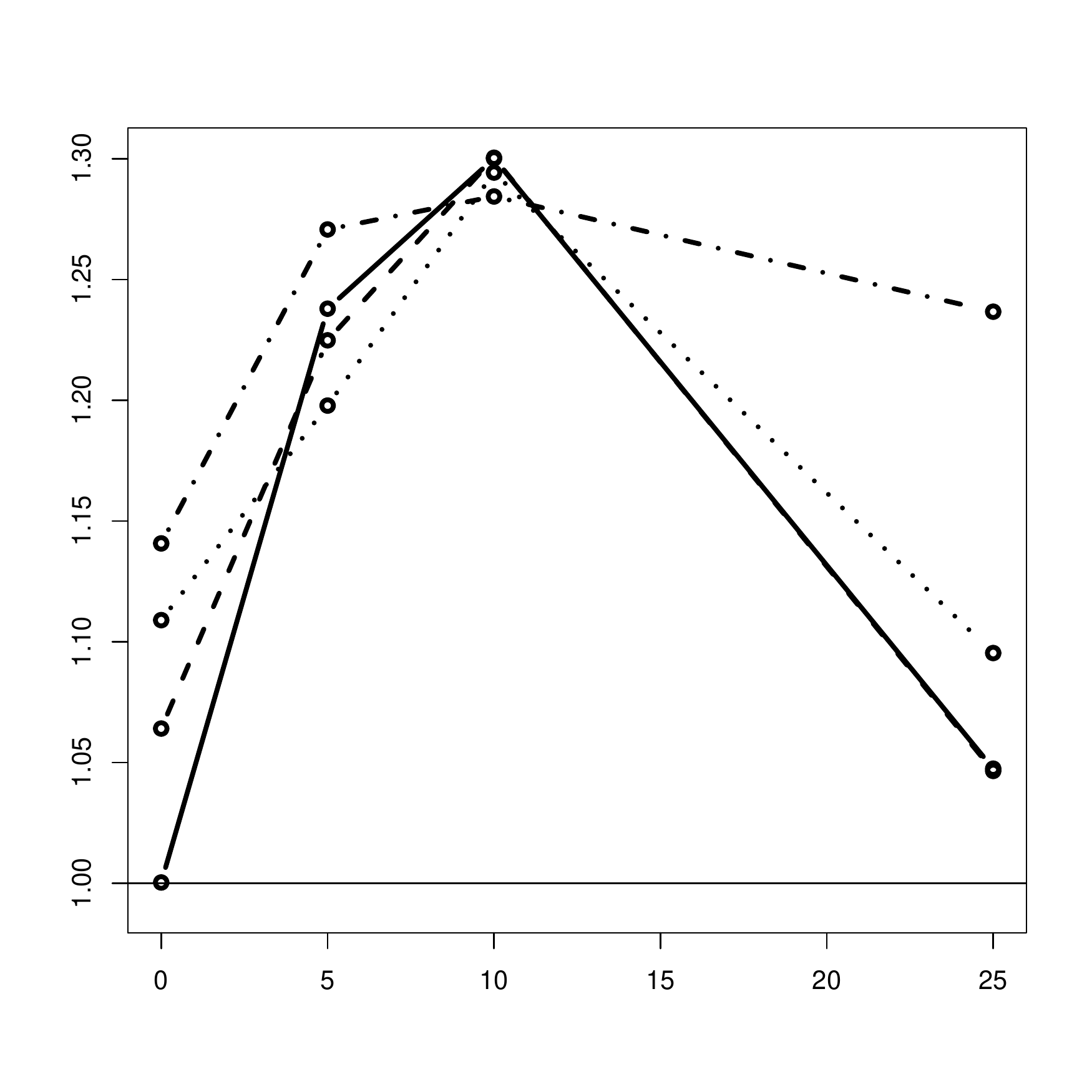}} \\
    \subfigure[$\beta_1$ under close contamination pattern ($f = 2$)]
     {\label{fig:eficbeta1f2}\includegraphics[width=0.45\linewidth]{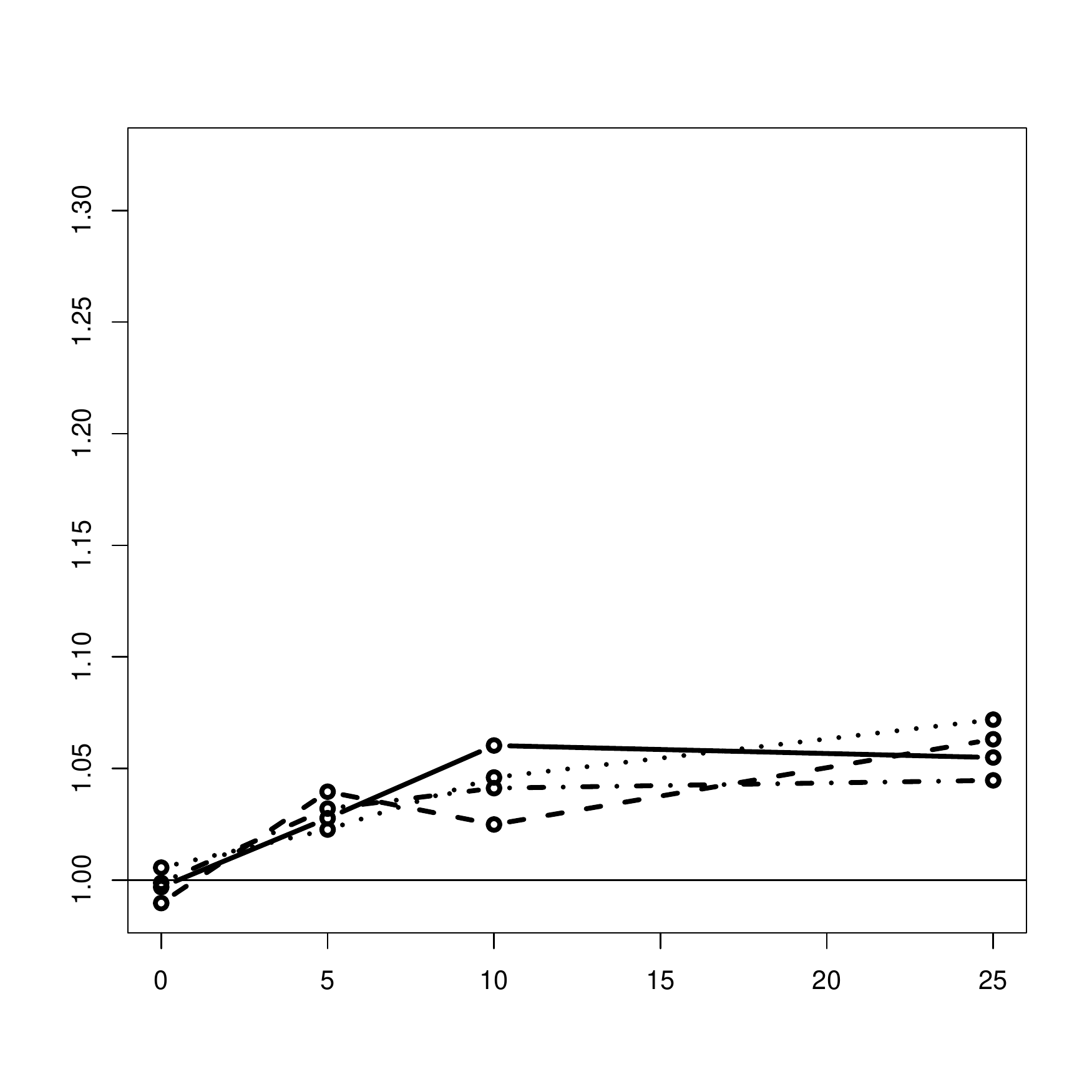}}
    \subfigure[$\beta_1$ under distant contamination pattern ($f = 4$)]
     {\label{fig:eficbeta1f4}\includegraphics[width=0.45\linewidth]{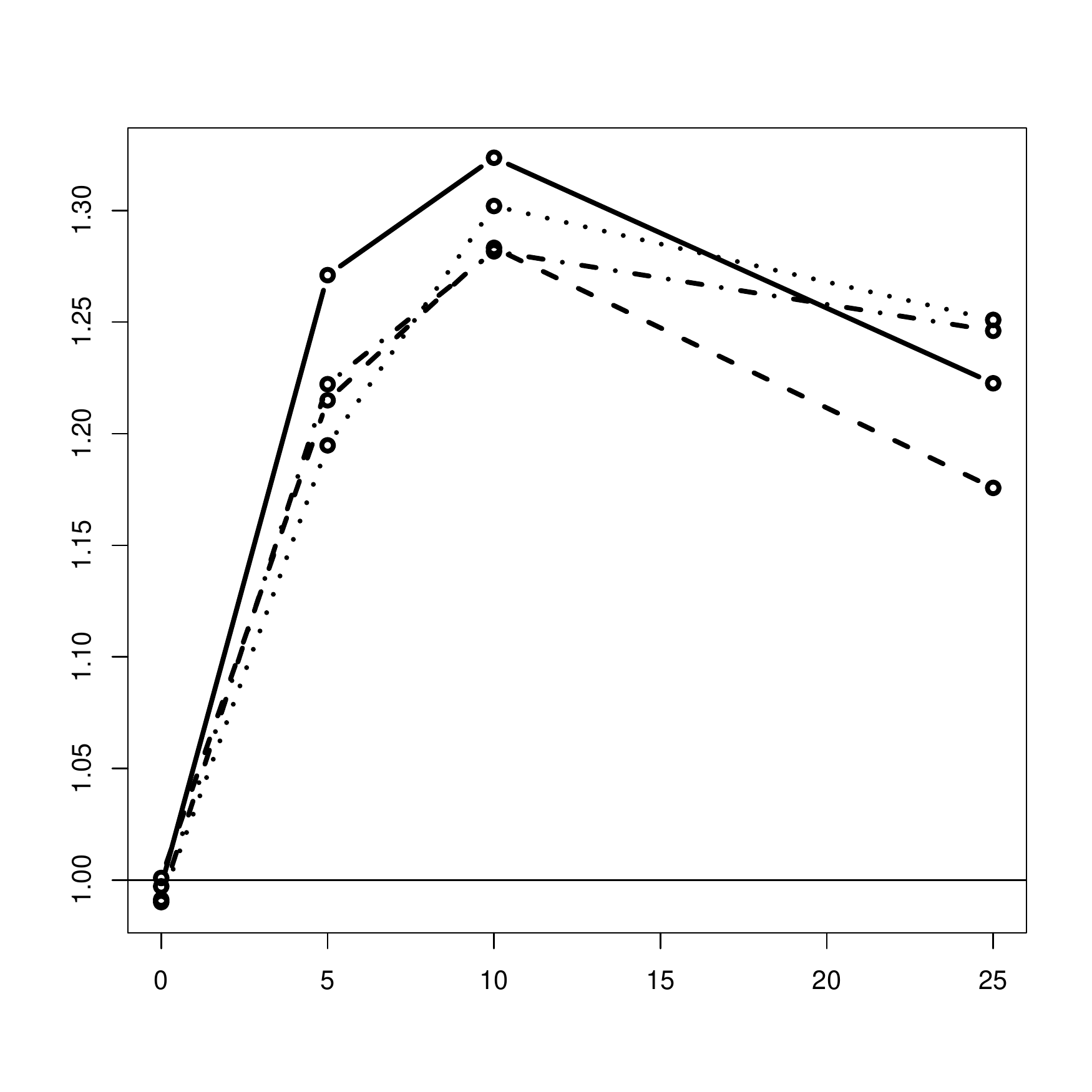}}
  \caption{Relative efficiency (see Eq. (\ref{eq:tmm-effic})) of the $t$ posterior mean with respect to the Gaussian 
posterior mean for the fixed effects in the linear mixed effect example. It plots the efficiency on the 
y-axis and $p_e$ on the x-axis. The meaning for the different types of lines are: (Solid line) $p_b = 0\%$, (Dashed line) $p_b = 5\%$, 
(Dotted line) $p_b = 10\%$ and (Dot-sash line) $p_b = 25\%$.}
\label{fig:cont1}
\end{figure}

\begin{figure}[ht!]
  \centering
    \subfigure[$\tau_e = 1/\sigma_e^2$ under close contamination pattern ($f = 2$)]
     {\label{fig:efictauf2}\includegraphics[width=0.45\linewidth]{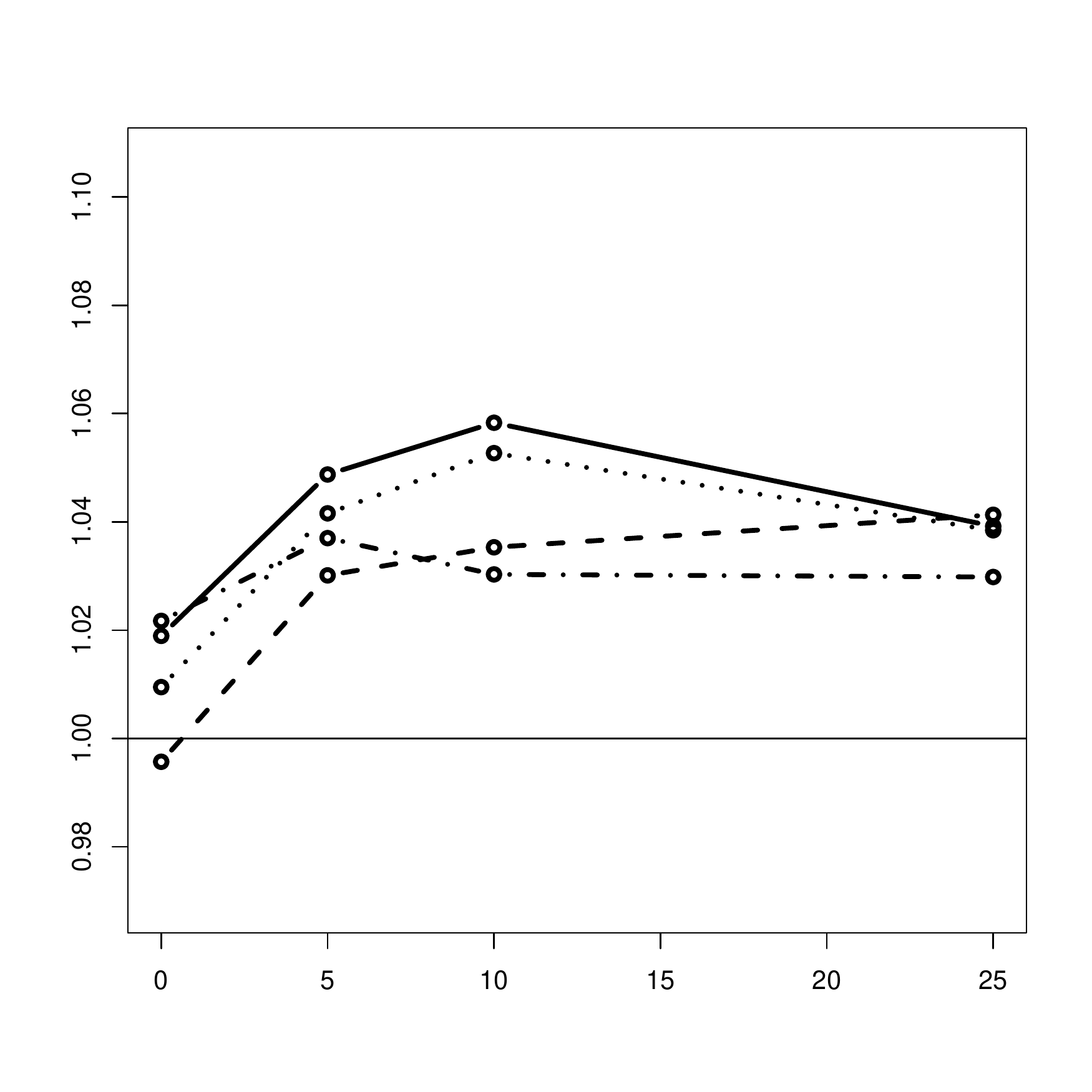}}
    \subfigure[$\tau_e = 1/\sigma_e^2$ under distant contamination pattern ($f = 4$)]
     {\label{fig:efictauf4}\includegraphics[width=0.45\linewidth]{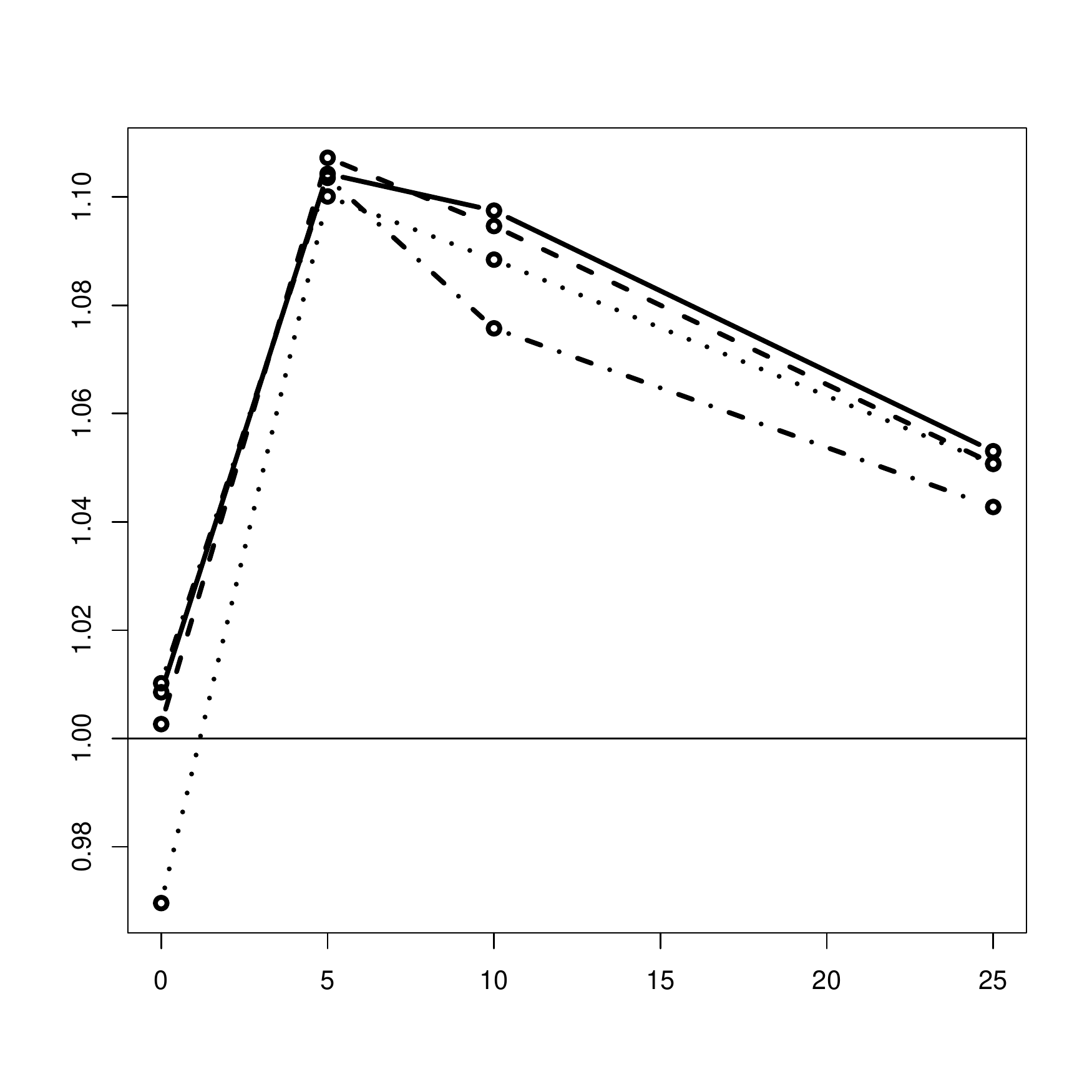}} \\
    \subfigure[$\tau_b = 1/\sigma_b^2$ under close contamination pattern ($f = 2$)]
     {\label{fig:efictaubf2}\includegraphics[width=0.45\linewidth]{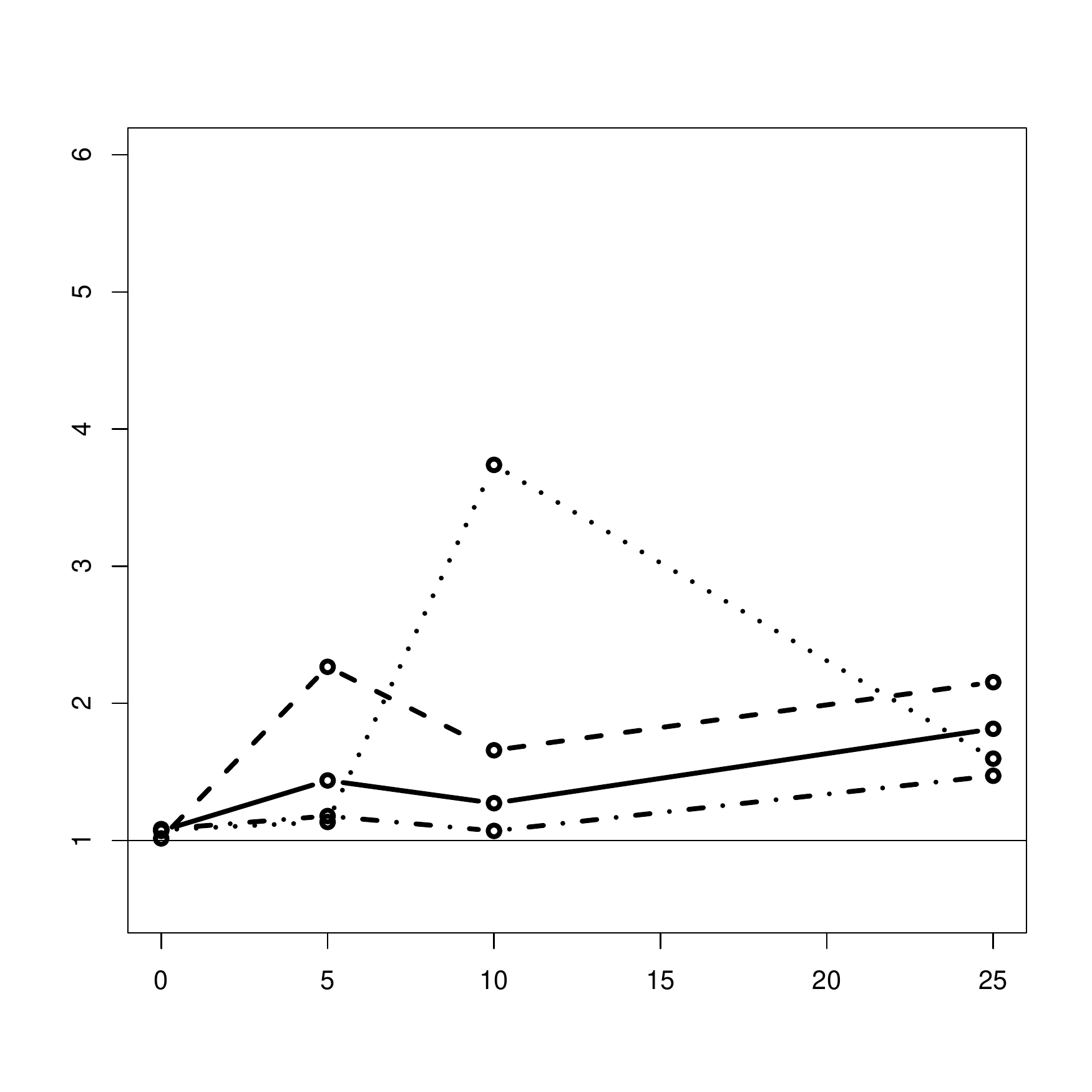}}
    \subfigure[$\tau_b = 1/\sigma_b^2$ under distant contamination pattern ($f = 4$)]
     {\label{fig:efictaubf4}\includegraphics[width=0.45\linewidth]{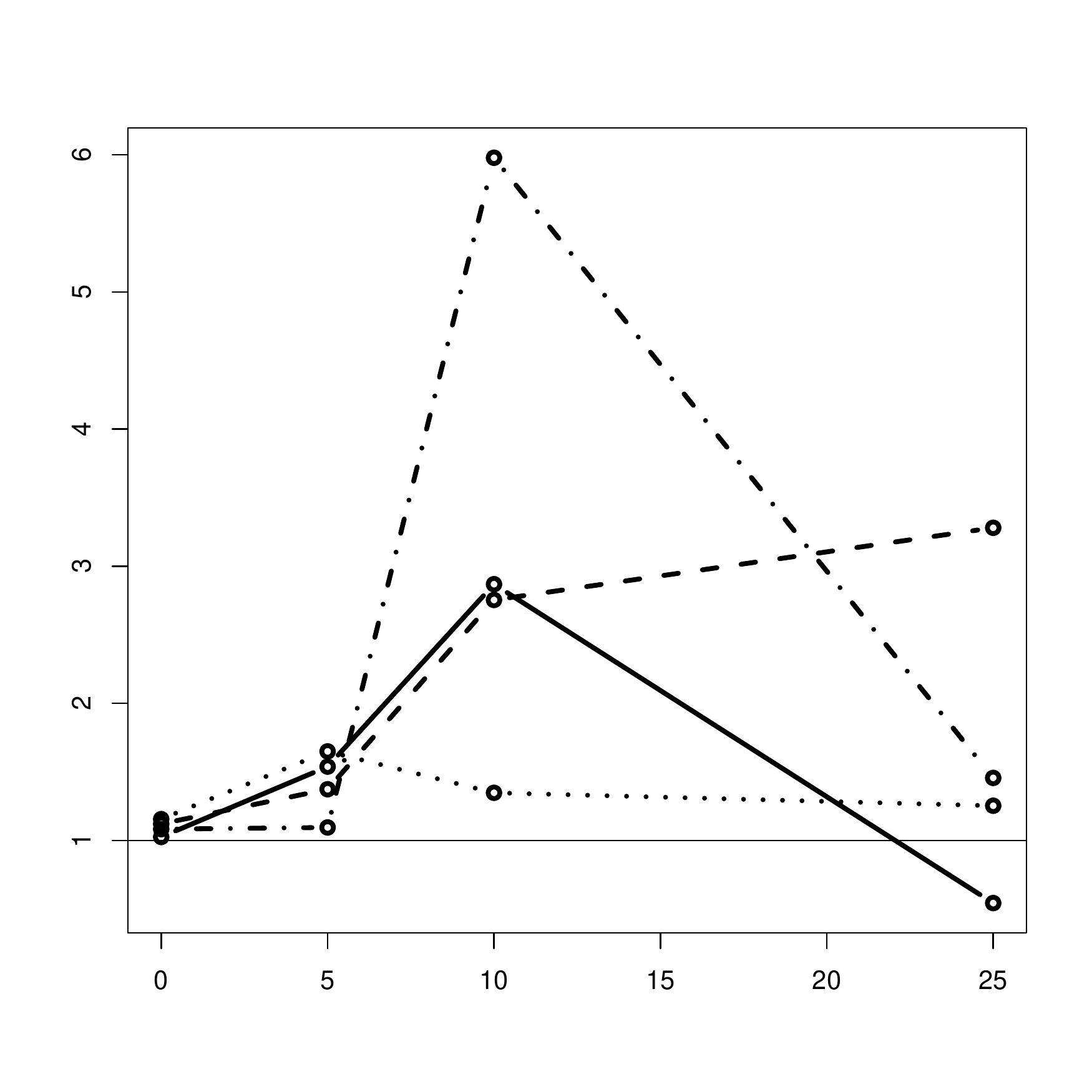}}
  \caption{Relative efficiency (see Eq. (\ref{eq:tmm-effic})) of the $t$ posterior mean with respect to the Gaussian 
posterior mean for the precision parameters in the linear mixed effect example. It plots the efficiency on the 
y-axis and $p_e$ on the x-axis. The meaning for the different types of lines are: (Solid line) $p_b = 0\%$, (Dashed line) $p_b = 5\%$, 
(Dotted line) $p_b = 10\%$ and (Dot-sash line) $p_b = 25\%$.}
\label{fig:cont2}
\end{figure}

Figures \ref{fig:cont1} and \ref{fig:cont2} plots the relative efficiency, defined in Eq. (\ref{eq:tmm-effic}),
between the posterior means of the $t$ model over the Gaussian linear mixed effects model. Based on the plots the 
conclusion of our simulation study are, as expected, similar to that obtained by \cite{pinheiro2001efficient}. There are 
substantial gains in efficiency for all parameters under the more distant contamination pattern ($f = 4$)
and moderate gains under the close contamination pattern ($f=2$). The efficiency gains are bigger for the 
precision of the random effects and the non-monotonic behavior of the efficiency gains suggest that the $t$ model 
is more robust than the Gaussian model especially for moderate percentage ($5-10\%$) of outliers. The two
methods have about the same efficiency under the no-contamination case.

\section{Conclusion}\label{sec:concl}

INLA is a deterministic approach to 
perform approximate fully Bayesian inference on the class of LGMs.
It is not meant to be a replacement of MCMC in general, but it is shown
to work extremely well in the broad class of LGMs, where it delivers 
accurate results with only a small fraction of computational time, when
compared to MCMC. Although many standard models currently in use by the
applied community fall within the class of LGMs, it is often necessary
to build more flexible models that go beyond the Gaussian distribution
for some of the latent components. This paper
describes the INLA extension that allow us to fit these more flexible models within
the INLA framework. The 
main idea is to approximate the non-Gaussian components of the latent field with a
Gaussian distribution and correct this approximation with a correction term in the 
likelihood. This solution preserves the Gaussian nature of the latent field, which is
of extreme importance for the INLA method. At the same time, it preserves the 
non-Gaussianity of the latent field in the model formulation. We have discussed 
the impact of this new formulation within the INLA methodology. Also, computational
considerations regarding the use of INLA in this more complex models were described.
Our approach works well for distributions that add flexibility around the Gaussian by correcting it in terms of skewness and/or kurtosis, which we denote as near-Gaussian distributions.

Two examples of interest were given. The first, survival model with gamma frailty, 
has log-concave likelihood and correction term, and is therefore considered 
easier from an optimization point of view. The second, a linear mixed model with 
Student's t distribution for the error term and for the random effects, is more challenging 
since both the likelihood and the correction term are not log-concave. Our approach
has provided very accurate approximations for posterior marginals and summaries 
for both examples when compared with very long MCMC runs. The comparison with MCMC
is made only for illustrative purposes, since the whole point of our approach 
is to avoid the use of time consuming MCMC algorithms. One nice property of our
approach is that the same techniques described in \cite{rue2009approximate} to 
check the accuracy of the the approximations can still be used. A summary of these
techniques were presented. 

Our extension 
can be used to fit non-Gaussian state space models 
\citep{kitagawa1987non} where the distribution of the noise in the state space evolution
equation is assumed to be non-Gaussian and will be discussed elsewhere.
There is also a plan to extend the R package \vb{INLA} to include options that allow the user to add flexibility and challenge the Gaussian assumptions of some components of the 
latent field in a straightforward and intuitive way. The focus of this extension will be on near-Gaussian distributions, and the method presented here will be used as the inference 
tool.

\appendix

\section{R code - Survival model}\label{app:survcode} 

Following is the INLA code used in example \ref{sec:examples:surv}. 
The code below show how we apply our idea of approximating the
non-Gaussian components of the latent field by a Gaussian distribution and
correcting this approximation with a correction term in the likelihood.
The faked observations is what allow us to include the correction term
in the likelihood. Notice that the \vb{family} argument in the \vb{inla}
function is what determine what is the ``likelihood" of this faked observations,
which in this case is the correction term in Eq. (\ref{eq:ex:multisurv:logCTi}),
defined as \vb{"loggammafrailty"} in \vb{R-INLA}. The expression 
\begin{verbatim}
f(loggamma.frailty, model="iid", 
                    hyper = list(prec = list(initial=-5, fixed=TRUE)))
\end{verbatim}
in the model formula is what indicates that the non-Gaussian components
of the latent field are being approximated by a Gaussian with mean zero
and low precision, $\exp(-5)$ in this case.
Please
visit \url{http://www.r-inla.org/} for more information about the R package
INLA.

\small{
\begin{verbatim}

# Simulate a dataset
#--------------------
n = 100 # number of groups
m = 10  # number of individuals in the same group
z = runif(n*m) # simulate covariate
eta = 1 + z    # linear predictor
frailty = rgamma(n, 1, 1) # simulate frailties
y = rexp(n*m, rate = rep(frailty, each = m) * exp(eta)) # simulate data

# INLA code
#------------
## Construct an extended response vector Y.
yy = inla.surv(c(y, rep(NA, n)), 
               c(rep(1, n*m), rep(NA, n))) # Observation component
ff = c(rep(NA, n*m), rep(1, n)) # Frailty component, 
                                # any observation will do, like '1'
Y = list(yy, ff) # extended response vector


## Construct extended covariates and frailties
intercept = c(rep(1, n*m), rep(NA, n))       # intercept
zz = c(z,  rep(NA, n))                       # covariate
loggamma.frailty = c(rep(1:n, each=m), 1:n)  # frailty
## Model formula
formula = Y ~ -1 + intercept + zz + 
              f(loggamma.frailty, model="iid", 
                hyper = list(prec = list(initial=-5, fixed=TRUE)))
## prior for the frailty
hyper.frailty = list(prec = list(param=c(1, 1)))
## Run inla function
rr = inla(formula,  
          data = list(Y=Y, zz=zz, intercept=intercept, 
                      loggamma.frailty= loggamma.frailty),
          family = c("exponential",  "loggammafrailty"),
          control.data = list(list(),  list(hyper = hyper.frailty)), 
          control.fixed = list(prec = list(default = 0.01)),
          control.inla = list(strategy = "laplace")
          )
\end{verbatim}
}

\bibliography{bibliografia_all.bib}

\end{document}